\documentclass[MS]{msu-thesis}
\usepackage[T1]{fontenc}
\usepackage{newtxtext,newtxmath} 
\usepackage{graphicx} 
\usepackage{xspace}

\usepackage{algorithm}

\usepackage{amsthm}
\usepackage{amsmath}
\usepackage{setspace}
\usepackage{subcaption}
\usepackage{listings}
\usepackage{color}

\newcommand{\repcl}[0]{\textit{RepCl}}
\newcommand{\vc}[0]{\textit{VC}}
\newcommand{\hlc}[0]{\textit{HLC}}

\newcommand{\clockskew}[0]{\mathcal{E}}
\newcommand{\epsilonone}[0]{\clockskew_1}
\newcommand{\epsilontwo}[0]{\clockskew_2}

\newcommand{\maxt}{mx\xspace}
\newcommand{\newmaxt}{newmx\xspace}

\newcommand{\intervalsize}[0]{\textit{I}}
\newcommand{\maxoffsetsize}[0]{\tau}
\newcommand{\offsetsize}[0]{\theta}
\newcommand{\countersize}[0]{\sigma}

\newcommand{\hb}[0]{\ \textit{hb}\ }

\newcommand{\br}[1]{\ensuremath{\langle #1 \rangle}\xspace}

\newcommand{\maxph}{mxph\xspace}

\definecolor{dkgreen}{rgb}{0,0.6,0}
\definecolor{gray}{rgb}{0.5,0.5,0.5}
\definecolor{mauve}{rgb}{0.58,0,0.82}

\title{Tracing Distributed Algorithms Using Replay Clocks}
\author{Ishaan Lagwankar}
\fieldofstudy{Computer Science and Engineering} 
\date{2024}

\begin{document}

\frontmatter
\maketitlepage

\begin{abstract}
 In this thesis, we introduce replay clocks ($\repcl$), a novel clock infrastructure that allows us to do offline analyses of distributed computations. The replay clock structure provides a methodology to \textit{replay} a computation as it happened, with the ability to represent concurrent events effectively. It builds on the structures introduced by vector clocks ($\vc$) and the Hybrid Logical Clock ($\hlc$), combining their infrastructures to provide efficient replay. With such a clock, a user can replay a computation whilst considering multiple paths of executions, and check for constraint violations and properties that potential pathways could take, especially in the presence of concurrent events. Specifically, if event $e$ must occur before $f$ then the replay clock must ensure that $e$ is replayed before $f$. On the other hand, if $e$ and $f$ could occur in any order, replay should not force an order between them. 
 
 After identifying the limitations of existing clocks to provide the replay primitive, we present the $\repcl$ structure and identify an efficient representation for the same. We demonstrate that $\repcl$ can be implemented with less than four integers for 64 processes for various system parameters if clocks are synchronized within $1ms$.
 
 Furthermore, the overhead of $\repcl$ (for computing/comparing timestamps and message size) is proportional to the size of the clock. Using simulations in a custom distributed system and NS-3, a state-of-the-art network simulator, we identify the expected overhead of $\repcl$ based on the given system settings. We also identify how a user can then identify feasibility region for $\repcl$. Specifically, given the desired overhead of $\repcl$, it identifies the region where unabridged replay is possible.  
 
 Using the $\repcl$, we provide a tracer for distributed computations, that allows any computation using the $\repcl$ to be replayed efficiently. The visualization allows users to analyze specific properties and constraints in an online fashion, with the ability to consider concurrent paths independently. The visualization provides per-process views and an overarching view of the whole computation based on the time recorded by the $\repcl$ for each event.
\end{abstract}

\clearpage

\makecopyrightpage 

%
%
\clearpage
\chapter*{Acknowledgements}
\DoubleSpacing 
First and foremost, I would like to express my deepest appreciation to my committee in supporting me through this work in the course of my Master's program. I am deeply indebted to Dr. Sandeep S. Kulkarni, and the ideas he has put forth that allowed me to progress in this research. The completion of this thesis would not have been possible without his support. I am deeply indebted to Dr. Li Xiao and Dr. Philip K. McKinley for providing me guidance through the committee and providing feedback for the work done in this research.
I would like to extend my sincere thanks to the Department of Computer Science and Michigan State University, with gratitude to Vincent Mattison and Brenda Hodge, for supporting me with administrative decisions and financial support for the ICDCN '24 conference in which this work was first published. 
I would also like to thank my peers in the Department, for their constant support and feedback on the work I have done, without which this thesis would not be complete. 
Lastly, I express deep gratitude to my family for providing me with the opportunities that brought me to this stage, and I will be forever indebted to them for their constant nurturing and support throughout my academic career.
\clearpage
\SingleSpacing
\tableofcontents* 
%
%
%
\mainmatter
%

\chapter{Introduction}

According to the observer effect, when we try to measure something, we change it to some extent \cite{mytkowicz2008observer}. 
Therefore, precise measurement is never truly possible. Computer programs suffer from this same difficulty. When you try to measure something in a computation, it changes the underlying computation. In an ideal world, a program may want to make sure that every step that it is taking is correct with respect to any environmental changes. However, the time taken for performing these checks may cause the program to be incorrect. In other words, it is possible that adding excessive safety checks or checks for guaranteeing fairness may cause the system to spend substantial time on those checks, thereby violate system requirements, even though those requirements would never have been violated without those checks in the first place. 

This issue is even more complicated in distributed computing where each process (component, node, etc.) relies on partial information. Hence, computing the required safety checks (or checking for the satisfaction of fairness requirements, etc.) would require processes to communicate with each other. In turn, the time for computing them would be even higher. 

As an illustration, consider two drones $A$ and $B$ that are cooperating to perform a task. Each drone may take independent actions based on some environment that the other drone cannot see. It is required that the area covered by the drones remains 50\% or above at all times (100\% of the time). It is also preferred that this area remains at 75\% most of the time (75\% of the time). 
Here, we would like to know (1) how frequently a given point $x$ was covered by one of the drones, (2) how frequently a given point was covered by both the drones, (3) what was the minimum coverage at any time, etc. 
(Assume that they are at two different altitudes so safety measures such as preventing collision are not necessary). 
One possibility is to require $A$ to notify $B$ of its actions at all times to ensure that $B$ can adjust its plan to account for what $A$ is doing, and vice versa. However, this will require $A$ to unnecessarily spend more time communicating with $B$, and vice versa. In other words, it may be necessary for $A$ and $B$ to move independently. Doing all these checks during the execution would require that $A$ and $B$ communicate with each other before they make any move. In turn, it would change the behavior of the drones completely thereby preventing us from making any conclusion about their behavior in the absence of these checks. Furthermore, the problem would be even more complicated if we had a larger number of drones. 

One way to address this problem is to log the computation as it is happening so we can evaluate it later for all properties of interest.
These properties may include non-critical safety properties or desirable performance criteria, etc. To be beneficial, the amount of storage for the log or the time to create that log should be small enough so that the underlying computation is affected minimally.
In other words, the measurement should not change the underlying computation substantially. 
At the same time, the log should capture the non-determinism that is inherently present in any distributed computation.  
We also want to make sure that the creation of the logs is performed independently by each process, i.e., each process stores its local state whenever it changes along with a timestamp (discussed next) that identifies when the change was made. 
We also assume that all messages are logged as well. 
We consider various approaches for storing the timestamps and their implications.  

The simplest approach we can consider is to let the timestamp be the physical time of the relevant process. Here, the storage and computation cost is very low. However, the physical clocks of processes often differ. Hence, drone $A$ may send a message at time 50 (local time of $A$) but it is received by $B$ at time $40$ ($B$'s local time). When we try to reply to this log to evaluate the given properties, it will cause $B$ to receive the message before $A$ has sent it. This is unacceptable, as it violates the system's consistency.

The next approach we can consider is vector clocks. Vector clocks introduce two concerns: Their size of $O(n)$, where $n$ is the number of processes, may be too high. Another challenge is that vector clocks do not have any reference to the physical clock and do not account for communication outside the system. For example, it is possible that drone $A$ activated a green LED at physical time $t_1$ and $B$ activated a white LED at time $t_2$ where $t_2 >> t_1$. In other words, an external observer will know that the action of $A$ occurred before $B$. However, if $A$ and $B$ did not communicate then the corresponding events will be concurrent \cite{Lamport78CACM}. Thus, when we replay the log, it is possible that the white LED event could be replayed before the green LED event. This is also unacceptable. 

Hybrid logical clocks (HLC) \cite{kulkarni2014logical} combine logical clocks and physical clocks. Specifically, they rely on a system where physical clocks are synchronized within the acceptable limit of clock skew, $\clockskew$, they guarantee that $hlc.e < hlc.f$ if $e$ happened before $f$ or $pt.e+\clockskew < pt.f $ (\cite{Lamport78CACM}). Here, $hlc.e$ denotes the Hybrid Logical Clock of process $e$, and $pt.e$ denotes the physical time observed on process $e$. In other words, $hlc.e < hlc.f$ if $f$ causally depends upon $e$ or $f$ occurred substantially after $e$. 
They eliminate the problem associated with physical clocks as HLC respects causality. They also eliminate the problem caused by vector clocks as the HLC timestamp of activating the green LED will be less than the $HLC$ timestamp of activating the white LED. 
$HLC$ does create another problem though. Consider the case where we have events $e$ and $f$ such that $|pt.e - pt.f| < \clockskew$ and $e||f$, i.e., the events are causally concurrent and very close to each other in physical time.  Without loss of generality, let $hlc.e < hlc.f$. In this situation, when we replay the log, $e$ will always occur before $f$. In other words, the log does not have the necessary information that could allow it to replay $f$ before $e$ even though they could have occurred in any order.

An extension of HLC, hybrid vector clocks \cite{yingchareonthawornchai2018analysis} reduces some of these issues. However, as we highlight in Section \ref{sec:properties}, this overhead is still quite high. 

Based on these limitations, in this thesis, we focus on building a new clock, the Replay Clock ($\repcl$), that combines hybrid logical clocks and vector clocks to eliminate their limitations. Our goal is to investigate scenarios under which $\repcl$ permits efficient replay of events. To understand why we may need to limit $\repcl$ to specific scenarios, observe that if the underlying system was asynchronous (unbounded clock drift) then it is required to have $O(n)$ vector clocks to enable replay of events. Systems that communicate frequently will need more information stored to replay events. Thus, we focus on the following problem: \textit{Given the amount of permissible overhead for logging events, what are the scenarios where perfect replay of events is possible?} 

Once we identify the scenarios in which perfect replay is available, we design a trace visualizer, named \textit{RepViz}, that allows us to depict candidate traces with the ability to replay concurrent events in any order of execution. This visualizer takes in a $\repcl$-timestamped trace to generate a visualization depicted in Chapter \ref{chap:visualization}. We provide per-process views with timelines of events to depict the events occurring while indicating causality between those events. The trace visualizer provides a interactive display to the user with orderings of events, and allows the user to reorder concurrent events to view different candidate traces. The visualizer API is discussed in Chapter \ref{chap:visualization}. 


\section{Contributions}
\begin{itemize}
    \item We present $\repcl$, a replay clock that enables the replay of events in a distributed system. It guarantees that if there is a causal relation \cite{Lamport78CACM} or if $f$ occurred far later than $e$ then $\repcl.e < \repcl.f$, i.e., the replay will cause $e$ to replayed before $f$. On the other hand, if $e$ and $f$ are causally concurrent and occurred close in physical time then they could be replayed in any order.
    \item By considering various system parameters, clock skew ($\clockskew$), message rate $(\alpha)$, and message delay $(\delta)$, we identify the feasibility region for $\repcl$.
    \item We implement an API for the $\repcl$ for NS-3, a widely used distributed network simulator. It provides all the operations and documentation on how it integrates with different network components available to the NS-3 simulations.
    \item We design \textit{RepViz}, a visualizer that generates traces from $\repcl$-timestamped logs in a distributed computation. The visualizer provides the user with various orders of replay, and allows the user to view different candidate traces and evaluate various constraints along those traces through the visualization.
\end{itemize}

\paragraph{Organization of the thesis: } This thesis is organized as follows. In Chapter \ref{chap:preliminaries}, we describe the model of computation for distributed systems including the notion of causality and clock synchronization. We move forward to the idea of the replay clock and the problems it solved in Chapter \ref{chap:repcl}. in Chapter \ref{chap:algorithm} we describe the algorithms associated with the $\repcl$, and describe the properties of the $\repcl$. Additionally, we describe the method of representation for the $\repcl$ and describe the various overheads that are characteristic of the design of the clock. Chapter \ref{chap:setup} talks about the design of the simulators we used to collect metrics for the $\repcl$. These metrics are discussed in Chapter \ref{chap:simulation} with analysis on the size of the clock and feasibility of implementation of the clock. Chapter \ref{chap:visualization} talks about the design of the \textit{RepViz}, the visualization system for trace building using the $\repcl$. Chapter \ref{chap:related} discusses related work and identifies questions raised by $\repcl$. Finally, in Chapter \ref{chap:conclusion}, we conclude and discuss future work. 

\chapter{Preliminaries}
\label{chap:preliminaries}

A distributed system is a set of processes $1..n$. Each process has three types of events (1) $send$, where it sends a message to another process, (2) $receive$, where it receives a message from another process, and (3) $local$, where it performs some local computation. 

We define the happened-before (denoted by $\hb$) relation \cite{Lamport78CACM} among the events in a distributed computation. 
\begin{itemize}
    \item If $e$ and $f$ happened on the same process and $e$ occurred before $f$  then $e \hb f$.
    \item If $e$ was a send event and $f$ was the corresponding receive event  then $e \hb f$.
    \item The $\hb$ relation is transitive, i.e., if there exist events $e, f$, and $g$ such that $e \hb g$ and $g \hb f$ then $e \hb f$
\end{itemize}

We say that $e || f$ iff $\neg (e \hb f) \wedge \neg(f \hb e)$. In other words, $e$ is concurrent with $f$ iff $e$ did not happen before $f$ and $f$ did not happen before $e$. 

A timestamping algorithm assigns a timestamp for every event $e$ in the system as soon as the event is created. Additionally, the timestamping algorithm defines a $<$ relation that identifies how two timestamps are compared. 

As an illustration, Lamport's logical clock \cite{Lamport78} assigns an integer timestamp $l.e$ to every event $e$. The $<$ relation for Lamport's logical clocks is the standard $<$ for integers. 
Likewise, the physical timestamping algorithm assigns $pt.e$ for every event $e$ where $pt.e$ is the physical time of the process where event $e$ occurred when it occurred, and the  $<$ relation is the same as that over integers.
A vector clock \cite{Fidge87}\cite{mattern1988virtual} assigns event $e$ a timestamp $vc.e$ where $vc.e$ is a vector that includes an entry $vc.e.j$ for every process $j$. The $<$ relation on two vector clocks $vc.e$ and $vc.f$ requires that each element in $vc.e$ is less than or equal to the corresponding element in $f$ and some element in $e$ is less than the corresponding element in $f$. In other words, $vc.e < vc.f$ iff $(\forall j :: vc.e.j \leq vc.f.j) \wedge (\exists j :: vc.e.j < vc.f.j)$.

Note that while the $<$ relation is defined by the timestamping algorithm, the properties of the $<$ relation vary. For example, logical clocks provide one-way causality information, i.e., $e \hb f \Rightarrow l.e < l.f$. Vector clocks provide two-way causality information, i.e., $e \hb f \Leftrightarrow vc.e < vc.f$. By contrast, (unsynchronized) physical clocks may not provide any guarantees. For example, it is possible that $(e \hb f)$ and $pt.e \not < pt.f$ are simultaneously true. 

We assume that each process $j$ in the system is associated with a physical clock, $pt.j$. Clocks of processes are synchronized with a protocol such as NTP \cite{mills1991internet} such that the clock of two processes differ by at most $\clockskew$, where $\clockskew$ is a parameter, i.e., $\forall j, k :: |pt.j -pt.k| \leq \clockskew$. We also assume that individual clocks are monotonically increasing. 
We assume that messages are delivered with a \textit{minimum message delay} of $\delta$. We do not assume maximum message delay; it could be $\infty$ if messages are permitted to be lost. Since our focus is on the replay of events, if a message is lost, it simply implies that the corresponding received message is never replayed. 

\chapter{Replay with Clocks}
\label{chap:repcl}

In this chapter, we focus on how clocks can be used to replay a given computation. We also discuss some of the limitations of using logical clocks and vector clocks in the replay process. Note that the goal of this chapter is only to illustrate the concept and the goals of the replay; it does not focus on developing an \textit{efficient} algorithm for the same.  

As discussed in the introduction, the goal of replay is to order the events so that we can evaluate various properties of interest.
To replay a given computation, we begin with the set where each entry is of the form $\br{e, ts.e}$, where $e$ is the event (send/receive/local) and $ts.e$ is the timestamp of $e$. To replay the given set of events, we first find events $e$ such that all events with smaller timestamps than $e$ have already been replayed. In other words, we find the set $\{e | \neg (\exists f: ts.f < ts.e)\}$. We replay one of these events randomly. Then, we remove event $e$. The process is continued until all events are replayed. Thus, the algorithm for replay is shown in algorithm \ref{alg:replay-events}.

\begin{algorithm}
\caption{ReplayEvents Operation}
\label{alg:replay-events}
\begin{enumerate}
    \item \textbf{Input:} $S$: Set of Events and timestamp
    \item \textbf{While} $S \neq \phi$ \textbf{do}
    \item \quad $FrontLine = \{ e | (e, ts.e) \in S \wedge \neg (\exists f: (f, ts.f) \in S \wedge ts.f < ts.e)\}$
    \item \quad Choose a random event $e$ from $FrontLine$ and replay it
    \item \quad $S = S - \{e\}$
    \item \textbf{end while}
\end{enumerate}
\end{algorithm}

\section{Limitations of Existing Clocks for Replay}
\label{sec:limitations}
\begin{figure}[ht]
    \centering
    \includegraphics[width=0.4\linewidth]{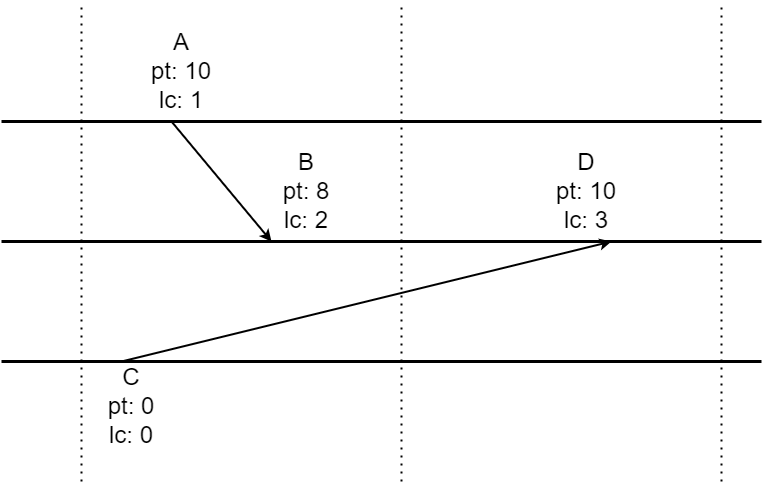}
    \caption{Sample Execution Sequence and Application of Replay Algorithm.}
    \label{fig:intro-example}
\end{figure}

As an example, consider the execution in Figure \ref{fig:intro-example}. Here, we have four events $A, B, C$, and $D$. Their physical timestamps and logical timestamps are shown in Figure \ref{fig:intro-example}. If we replay these events using physical clocks then the possible outcomes are $CBAD$ or $CBDA$. Note that both these outcomes are undesirable, as $A$ should occur before $B$ based on the causality (happened-before) relation. 

If we replay them by logical clocks, the possible outcome is $CABD$. However, there is no option to replay $B$ before $C$ even though $B||C$. 

If we replay them with vector clocks, possible orderings are $ABCD$, $ACBD$, or $CABD$. If the clocks were synchronized to be within 5 time units then $ABCD$ and $ACBD$ are incorrect.

\section{Requirements of Replay Clock ($\repcl$)}
\label{sec:newrepclrequirements}

In this thesis, we focus on a system where the physical clocks are synchronized to be within $\clockskew$, i.e., for any two processes $j$ and $k$, $|pt.j-pt.k| \leq \clockskew$. The goal of $\repcl$ is to assign a timestamp $\repcl.e$ to event $e$ such that 

\begin{description}
    \item [Requirement 1. ] If $e$ happened before $f$ then\\ $\repcl.e < \repcl.f$, i.e., $e$ will always be replayed before $f$
    
    \item [Requirement 2. ] If $f$ occurred far after $e$, i.e., $e$ and $f$ could not have occurred simultaneously under clock drift guarantee of $\epsilonone$
    where $\epsilonone 
    \approx 
    \clockskew$ then 
    $e$ will be replayed before $f$, i.e., $\repcl.e < \repcl.f$

    \item [Requirement 3. ] If $e$ and $f$ could have occurred in any order in a system where clocks were synchronized to be within $\epsilontwo$, where $\epsilontwo\approx\clockskew$ then $\repcl.e||\repcl.f$ (i.e., $\neg (\repcl.e < \repcl.f) \wedge \neg(\repcl.f < \repcl.e))$   

\end{description}

In the last two requirements, we have chosen $\epsilonone$ and $\epsilontwo$ instead of $\clockskew$ itself as it can permit more efficient implementation by allowing us to maintain a coarse-grained clock. We discuss this further in section \ref{sec:properties}.

With these requirements in mind, we show that the $\repcl$ provides efficient replays of distributed computations, without suffering with the overhead that vector clocks impose, and the shortcomings of replay with logical clocks like the HLC. The $\repcl$ combines the best of both these worlds, and provides a better mechanism to replay computations.

\chapter{Algorithm for the Replay Clock ($\repcl$)}
\label{chap:algorithm}

In this chapter, we present our approach for $\repcl$. As discussed earlier, we assume that the physical clocks are synchronized to be within $\clockskew$. We discretize the process execution in terms of epochs, where each epoch corresponds to an increment of the clock by $\intervalsize$, $0 < \intervalsize \leq \clockskew$ such that $\clockskew = \epsilon*\intervalsize$, where $\epsilon$ is an integer. 
The timeline of a process is split into epochs where each epoch is of size \intervalsize (in the local process clock). In other words, the epoch of process $j$ is obtained by $\lfloor \frac{pt.j}{\intervalsize}\rfloor$.

\section{Structure of $\repcl$ timestamp. }
With such discretization, the timestamp of process $j$ (or event $e$) is of the form 
\begin{equation}
    \br{\maxt.j, \texttt{bitmap}.j[], \texttt{offset}.j[], \texttt{counter}.j[]},
\end{equation}
where $\maxt.j$ is an integer for the approximation of the top-level $HLC$, and $\texttt{bitmap}.j$, $\texttt{offset}.j$ and $\texttt{counter}.j$ are bitsets \cite{schildt1998c++} that store \textit{at most} one entry $\texttt{bitmap}.j.k$, $\texttt{offset}.j.k$ and $\texttt{counter}.j.k$ for process $k$. Each of these bitsets is treated as an array but are serialized as integers in packets for efficiency. 

The intuition behind these variables is as follows:
\begin{itemize}
    \item $\maxt.j$ denotes the maximum epoch process $j$ is aware of (either due to the value of $pt.j$ or the value of epochs learned from messages it receives). 
    \item $\texttt{bitmap}.j.k$ is essentially an array of bits, where each bit with index $k$ denotes whether the $\texttt{offset}.j.k$ is being stored. If the bit is 1, process $j$ is actively maintaining $\texttt{offset}.j.k$. This will come in handy for efficient updates to the clock.
    \item $\maxt.j-\texttt{offset}.j.k$ denotes the maximum epoch value of $k$ that $j$ has learnt (either via direct/indirect message from $k$, clock drift assumption, etc). If there exists an offset between two processes $j$ and $k$, $\texttt{offset}.j.k$ denotes the difference between $\maxt.j$ and $\maxt.k$ as seen on process $j$.
    \item Counters are used to deal with the scenario where multiple events happen within the same epoch, and have the same offsets. If two clocks that are not concurrent have the same $\maxt$ and the same $\texttt{offset}$ values, then the two clocks differ on the counters. The clock with the lower counter value is replayed first.

\end{itemize}

For example, the timestamp $\br{50, [1, 1, 1], [0, 1, 2],[4, 5, 6]}$ denotes that this event is aware of epoch 50 of process $0$ (as $50 - 0$), 49 of process $1$ (as $50-1$), and 48 of process $2$ (as $50 -2$). And, the counter values are $4, 5$ and $6$ respectively. 

\section{Efficient traversal and lookup}

All computations are optimized using the bitmap. While the bitmap does not serve a purpose to the timestamp itself, it allows us to traverse and update the clock efficiently. Here we describe the traversal and lookup of offsets based on the bitmap. For brevity, we describe the rest of the algorithms as a simple traversal, but it is important to note that each traversal takes O(number of 1s in the bitmap) time complexity, and getters and setters take O(1) complexity. To describe these implementations, we use the integer representations of $\texttt{offset}.j[]$ and $\texttt{counter}.j[]$.

\subsection{Traversal}

The traversal operation is described in algorithm \ref{alg:traverse}, which iterates through the $\texttt{bitmap}$ to find all processes for which the $\texttt{offset}$ is being maintained. In the algorithm, we get every index that has a set bit in the bitmap. The set bit at position $k$ indicates that $\texttt{offset}.j.k$ is being stored by process $j$.

\begin{algorithm}
\caption{Traversal Operation}
\label{alg:traverse}
\begin{enumerate}
    \item Define Traversal operation.
    \item Traversal(${ts}$)
    \item \textbf{While} $ts.bitmap > 0$
    \item \quad $index = \log_2 (\neg (ts.bitmap \oplus (\neg (ts.bitmap - 1)) + 1) >> 1)$
    \item \quad // Get or set any offset at index
    \item \quad $ts.bitmap = ts.bitmap \land (ts.bitmap - 1)$
    \item \textbf{End While}
\end{enumerate}
\end{algorithm}

\subsection{Extract}

The extract operation extracts $k$ bits from position $p$. The algorithm is described in algorithm \ref{alg:extract}.

\begin{algorithm}
\caption{Extract Operation}
\label{alg:extract}
\begin{enumerate}
    \item Define Extract operation.
    \item Extract($number, k, p$)
    \item \textbf{Return} $((1 << k) - 1 \land (number >> p))$
\end{enumerate}
\end{algorithm}

\subsection{GetOffsetAtIndex}

GetOffsetAtIndex is an operation that gets the offset stored at a particular index. This is an O(1) lookup operation using the index obtained from the bitmap traversal or a specific offset that the clock may need at any index. The algorithm is described in algorithm \ref{alg:getoffset}. Here, $\maxoffsetsize$ denotes the max offset size allowed by the user (in bits). The max offset size is usually set to $log(\epsilon)$.

\begin{algorithm}
\caption{GetOffsetAtIndex Operation}
\label{alg:getoffset}
\begin{enumerate}
    \item Define GetOffsetAtIndex operation.
    \item GetOffsetAtIndex(${ts}, index$)
    \item $offset = Extract(offset.j[].ToInteger(), \maxoffsetsize, \maxoffsetsize * index)$
    \item \textbf{Return } $offset$
\end{enumerate}
\end{algorithm}

\subsection{SetOffsetAtIndex}

SetOffsetAtIndex is an operation that sets the offset at a particular index. This is an O(1) setter operation using the index obtained from the bitmap traversal or a specific offset that the clock may need at any index, like the GetOffsetAtIndex algorithm. The algorithm is described in algorithm \ref{alg:setoffset}. 

\begin{algorithm}
\caption{SetOffsetAtIndex Operation}
\label{alg:setoffset}
\begin{enumerate}
    \item Define SetOffsetAtIndex operation.
    \item SetOffsetAtIndex(${ts}, index, newoffset$)
    \item $firstpart = Extract(offsets.j, \maxoffsetsize*index, 0)$
    \item $res |= firstpart$
    \item $res |= newoffset << index * \maxoffsetsize$
    \item $lastpart = Extract(offsets.j, 
                            \maxoffsetsize * N - (\maxoffsetsize*(index + 1)),
                            \maxoffsetsize*(index + 1))$
    \item $res |= lastpart << (index + 1) * \maxoffsetsize$
    \item \textbf{Return } $res$
\end{enumerate}
\end{algorithm}

\subsection{RemoveOffsetAtIndex}

The RemoveOffsetAtIndex operation removes an offset given an index. This is an O(1) removal operation once the position is found by the traversal operation. This algorithm is described in Algorithm \ref{alg:remoffset}.

\begin{algorithm}
\caption{RemoveOffsetAtIndex Operation}
\label{alg:remoffset}
\begin{enumerate}
    \item Define RemoveOffsetAtIndex operation.
    \item RemoveOffsetAtIndex(${ts}, index, newoffset$)
    \item $firstpart = Extract(offsets.j, \maxoffsetsize*index, 0)$
    \item $res |= firstpart$
    \item $lastpart = Extract(offsets.j, 
                            \maxoffsetsize * N - (\maxoffsetsize*(index + 1)),
                            \maxoffsetsize*(index + 1))$
    \item $res |= lastpart << (index + 1) * \maxoffsetsize$
    \item \textbf{Return } $res$
\end{enumerate}
\end{algorithm}

\section{Helper functions}

Now that we have described the traversals and auxiliary operations, we move on to the clock helper algorithms. For brevity, we assume all traversals and assignments are the algorithms described in the previous subsection. We omit the bitmap to make it easier to understand the algorithms that follow. We discuss two helper functions, \textit{Shift} and \textit{MergeSameEpoch}, that will come in handy when we design the main clock processing algorithms.

\subsection{Shift Operation}

The Shift function allows us to change the value of $\maxt$. 
Since $\maxt.j-\texttt{offset}.j.k$ denotes the knowledge of $j$ has about the epoch of $k$, if $\maxt$ is changed to \newmaxt without providing $j$ any additional knowledge of the clock of $k$ then $\newmaxt-\texttt{newoffset}.j.k$ should remain the same as $\maxt.j - \texttt{offset}.j.k$. Hence, Shift operation changes offset$.j.k$ to be $\texttt{offset}.j.k+(\newmaxt-\maxt)$. Furthermore, if this value is more than $\epsilon$ then we reset it to $\epsilon$, as guaranteed by the clock drift assumption. (Note that process $j$ can learn about the clock of $k$ via clock synchronization assumption even if $j$ and $k$ do not communicate.)

\begin{figure}
    \centering
    \includegraphics[width=0.7\linewidth]{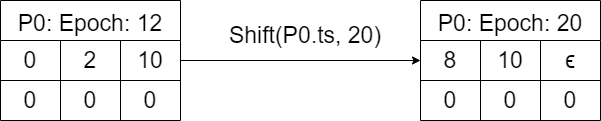}
    \caption{Working of Shift() on Process 0. Here, $\epsilon$ = 15, and the shift is issued to advance to $\maxt$ 20. Process 3's $\texttt{offset}$ becomes 20, but since $\epsilon$ = 15, the $\texttt{offset}$ is set to $\epsilon$ (due to clock skew limit guarantees).}
    \label{fig:shift}
\end{figure}

For example, shifting the timestamp $\br{12, [0, 2, 10]}$ so that $\maxt$ is changed to 20 will result in $\br{20, [8, 10,18]}$. If $\epsilon=15$, this will change to $\br{20, [8, 10, \epsilon]}$ (cf. Figure \ref{fig:shift}).

\begin{algorithm}
\caption{Shift Operation}
\label{alg:shift}
\begin{enumerate}
    \item Define Shift operation.
    \item Shift(${ts}, \newmaxt$)
    \item \textbf{For each} $k$ \textbf{do}
    \item \quad $ts.\texttt{offset}.k = {\texttt{offset}.k} + (\newmaxt - {ts.\maxt})$
    \item \quad \textbf{If} $ts.\texttt{offset}.k > \epsilon$ \textbf{then}
    \item \quad \quad $ts.\texttt{offset}.k = \epsilon$
    \item \quad \textbf{End If}
    \item \textbf{End For}
    \item \textbf{Output:} ${ts}$
\end{enumerate}
\end{algorithm}

\subsection{MergeSameEpoch Operation}

The MergeSameEpoch function takes two timestamps $t1$ and $t2$ with the same $\maxt$ value and combines their offsets by setting to be $\texttt{offset}.j.k$ to be the $min(t1.\texttt{offset}.j.k, t2.\texttt{offset}.j.k)$. For example, merging $\br{50, [0, 1, 2]}$ and $\br{50, [2, 0, 1]}$ results in $\br{50, [0, 0,1]}$.

\begin{algorithm}
\caption{MergeSameEpoch Operation}
\label{algo5}
\begin{enumerate}
    \item \textbf{Input:} Timestamp $t1$, Timestamp $t2$
    \item Timestamp $ts$ = new Timestamp
    \item \textbf{For each} $k$ \textbf{do}
    \item \quad $ts.\texttt{offset}.j.k = \min(t1.\texttt{offset}.j.k, t2.\texttt{offset}.j.k)$
    \item \textbf{End For}
    \item \textbf{Return} $ts$
\end{enumerate}
\end{algorithm}

\subsection{EqualOffset Operation}

The EqualOffset function takes in two timestamps $t1$ and $t2$ and checks whether the \texttt{offset} arrays and $\maxt$ values are the same. This is used particularly to update the \texttt{counter} array if the other values are equal. 

\begin{algorithm}
\caption{EqualOffset Operation}
\label{algo6}
\begin{enumerate}
    \item \textbf{Input:} Timestamp $t1$, Timestamp $t2$
    \item \textbf{If} $t1.{\maxt} \neq t2.{\maxt} \vee (\exists j \ t1.{\texttt{offset}.j} \neq t2.{\texttt{offset}.j})$ \textbf{then}
    \item \quad \textbf{Return} false
    \item \textbf{Else}
    \item \quad \textbf{Return} true
    \item \textbf{End If}
\end{enumerate}
\end{algorithm}

\section{Description of the $\repcl$ Algorithm} 

In this section, we discuss the key clock processing algorithms. These algorithms update the clock based on the type of event observed on the process. We describe two key operations: Send/Local and Receive. 

\subsection{Local/Send event} Here, we describe how $\repcl.j$ is updated when $j$ sends a message.
Let the current timestamp of $j$ be 
\begin{equation}
    \br{\maxt.j, \texttt{bitmap}.j[], \texttt{offset}.j[], \texttt{counter}.j[]}    
\end{equation}

First, $\maxt.j$ needs to be increased if the clock of $j$ has advanced beyond epoch $\maxt.j$. Hence, we first compute $\newmaxt.j$ which is equal to $max(\maxt.j, \texttt{epoch}.j)$. 
When $j$ sends a message, it does not learn any new information about the clock of process $k$. 

We consider two cases: The first case is for the scenario where the newly created event $f$ is in a new epoch as the previous event, $e$. This will happen if $\maxt$ remains unchanged and $\texttt{offset}.j.j$ is unchanged. In this case, we increase $\texttt{counter}.j.j$. 

The second case deals with the scenario where $f$ is in a new interval. Thus, the offset associated with $k$ is changed using the Shift operation. Note that the Shift operation computes the shift of all processes except $j$. $\texttt{offset}.j.j$ should be based on the value of $\texttt{epoch}.j$. Hence, we set it equal to $\newmaxt-\texttt{epoch}.j$. The Shift operation is illustrated in Algorithm \ref{alg:shift}.

\begin{algorithm}
\caption{Send Message}
\label{algo3}
\begin{enumerate}
    \item $\newmaxt = \max({\maxt.j}, {\texttt{pt}.j})$
    \item ${\texttt{new\_offset}} = {\newmaxt} - {\texttt{pt}.j}$
    
    \item \textbf{If} $({\maxt.j} = {\newmaxt} \land {\texttt{offset}.j.j} = {\texttt{new\_offset}})$ \textbf{then}
    \item \quad ${\texttt{counter}.j.j} = {\texttt{counter}.j.j} + 1$
    \item \textbf{Else}
    \item \quad ${ts.j} = {Shift}({ts.j}, {\newmaxt})$
    \item \quad ${\texttt{offset}.j.j} = \min({\newmaxt} - {\texttt{pt}.j},\epsilon)$
    \item \quad ${\texttt{counter}.j} = [0,0,\ldots,0]$
    \item \textbf{End If}
\end{enumerate}
\end{algorithm}

\subsection{Receive event}
Next, we describe how $\repcl$ is updated when $j$ with timestamp 
\begin{equation}
     \br{\maxt.j, \texttt{offset}.j[], \texttt{counter}.j[]}
\end{equation} 
receives a message $m$ with timestamp $\br{\maxt.m, \texttt{offset}.m[], \texttt{counter}.m[]}$.

First, we compute $\newmaxt$ which is the maximum of $\maxt.j$, $\maxt.m$ and $\texttt{pt}.j$. Timestamps of $j$ and $m$ are then shifted to $\newmaxt$ using the Shift operation. These timestamps are then merged to obtain the $\maxt$ and $\texttt{offset}$ values of the new event, say $f$. 

Now, we check if the knowledge that $f$ has about epochs is the same as that of $e$ (the previous event on $j$) or $m$. If all three are in the same epoch then $\texttt{counter}.j.j$ is set to one more than the maximum of $\texttt{counter}.j.k$ and $\texttt{counter}.m.k$. If only $e$ and $f$ are in the same epoch, $\texttt{counter}.j.j$ is incremented by $1$.
If only $m$ and $f$ are in the same epoch, $\texttt{counter}.j$ is set to $\texttt{counter}.m$ and the value of $\texttt{counter}.j.j$ is incremented by $1$. If none of these conditions apply then counters are reset to $0$. 

\begin{algorithm}
\caption{Receive Message}
\label{algo4}
\begin{enumerate}
    \item \textbf{Input:} Received Message $m$
    \item $\newmaxt = \max(\maxt.j, \maxt.m, \texttt{pt}.j)$
    \item ${ts.a} = {\texttt{Shift}}({ts.j}, \newmaxt)$
    \item ${ts.b} = {\texttt{Shift}}({ts.m}, \newmaxt)$
    \item ${ts.c} = {\texttt{MergeSameEpoch}}({ts.a}, {ts.b})$
    
    \item \textbf{If} $\texttt{EqualOffset}(ts.j, {ts.c}) \land$ $\texttt{EqualOffset}(ts.m, {ts.c})$ \textbf{then}
    \item \quad \textbf{For each} $k$ \textbf{do}
    \item \quad \quad $\texttt{counter}.j.k = \max(\texttt{counter}.j.k, \texttt{counter}.m.k)$
    \item \quad \textbf{End For}
    \item \quad $\texttt{counter}.j.j = \texttt{counter}.j.j + 1$
    \item \textbf{End If}
    
    \item \textbf{If} $\texttt{EqualOffset}(ts.j, {ts.c}) \land \neg$ $\texttt{EqualOffset}(ts.m, {ts.c})$ \textbf{then}
    \item \quad $\texttt{counter}.j.j = \texttt{counter}.j.j + 1$
    
    \item \textbf{If} $\neg$ $\texttt{EqualOffset}(ts.j, {ts.c}) \land$ $\texttt{EqualOffset}(ts.m, {ts.c})$ \textbf{then}
    \item \quad $\texttt{counter}.j = {\texttt{counter}.m}$
    \item \quad $\texttt{counter}.j.j = \texttt{counter}.j.j + 1$
    
    \item \textbf{If} $\neg$ $\texttt{EqualOffset}(ts.j, {ts.c}) \land \neg$ $\texttt{EqualOffset}(ts.m, {ts.c})$ \textbf{then}
    \item \quad $\texttt{counter}.j = [0, 0, \ldots, 0]$
\end{enumerate}
\end{algorithm}

\section{Comparing $\repcl$ Timestamps}
\label{sec:compare}

The \textit{happens-before} relation in $\repcl$ is codified in Algorithm \ref{alg:sort_events}. In this algorithm, we check whether timestamp $t1$ happens-before timestamp $t2$. In this relation, we first compare the HLCs of the two $\repcl$ timestamps. Since HLC provides the top-level information of the physical clock of the message, it resolves ties with clocks having different HLCs. If the HLCs are equal we move to the offsets. For $t1$ to be strictly happening before $t2$, we follow the comparison used in traditional vector clocks, where $vc.e < vc.f$ iff $(\forall j :: vc.e.j \leq vc.f.j) \wedge (\exists j :: vc.e.j < vc.f.j)$. If the offsets for the two timestamps are also equal, we check for the counters. 

We say two events are concurrent by the definition introduced in Requirement 3, which basically states that if $\neg(\repcl.e$ < $\repcl.f) \land \neg(\repcl.f$ < $\repcl.e)$, then $\repcl.e || \repcl.f$.

\begin{algorithm}
\caption{Compare Operation}
\label{alg:sort_events}
\begin{enumerate}
    \item \textbf{Input:} Timestamp $t1$, Timestamp $t2$
    \item \quad \textbf{If} t1.$\maxt$ < t2.$\maxt$ \textbf{then}
    \item \quad \quad \textbf{Return} $True$
    \item \quad \textbf{Else If} t1.$\maxt$ > t2.$\maxt$ \textbf{then}
    \item \quad \quad \textbf{Return} $False$
    \item \quad \textbf{Else} \textbf{then}
    \item \quad \quad \textbf{For} $i,j$ in t1.offsets, t2.offsets
    \item \quad \quad \quad \textbf{If} t1.offsets.$i$ > t2.offsets.$j$
    \item \quad \quad \quad \textbf{Return} $False$
    \item \quad \quad \textbf{If} t1.counters <= t2.counters
    \item \quad \quad \quad \textbf{Return} $True$
    \item \quad \textbf{End For}
    \item \quad \quad \textbf{Return} $False$
    \item \quad \textbf{End If}
\end{enumerate}
\end{algorithm}

\begin{figure}[ht]
    \centering
    \includegraphics[width=0.4\linewidth]{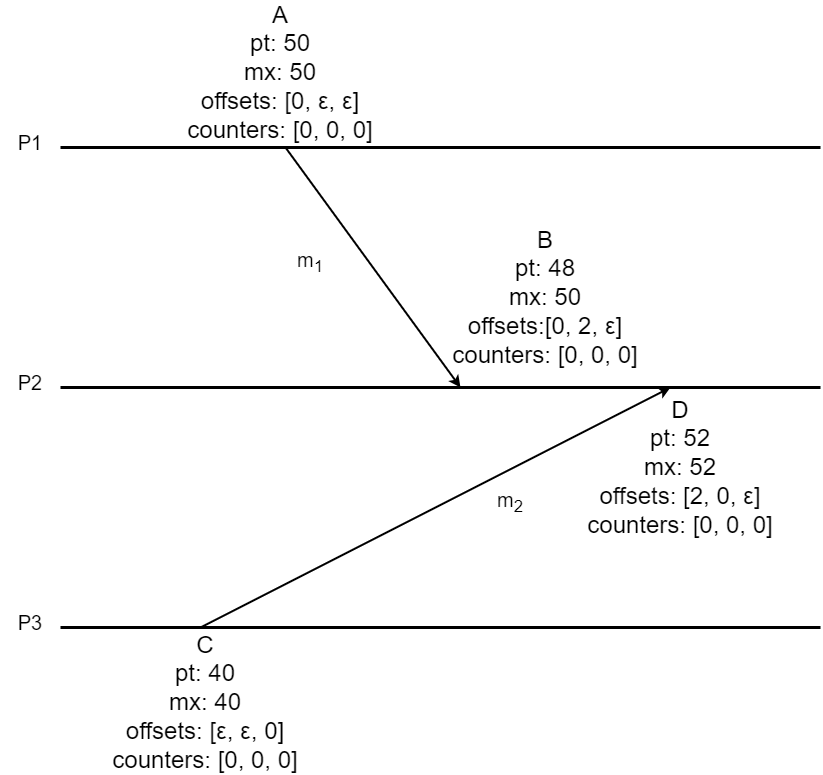}
    \caption{Replay of the Execution in Figure \ref{fig:intro-example} with $\repcl$.}
    \label{fig:repcl-example}
\end{figure}

As an illustration, consider the execution of the program in Figure \ref{fig:intro-example}. Assuming that $\epsilon=5,\intervalsize=1,$ and $\clockskew=5$, the $\repcl$ timestamps will be as shown in Figure \ref{fig:repcl-example}.
Here, event $A$ has physical time of $50$. Since process P1 has not heard from anyone else so far, the offsets for P2 and P3 will be $\epsilon$. The offset for process P1 will be $0$.
Regarding event $C$, the situation is similar except that the offset for process P3 is $0$. 
When event $B$ is created upon receiving message $m_1$, process P2 is aware of times $50$ from $P1$. And, it is the maximum epoch it is aware of. Hence, offsets are $[0, 2,\epsilon]$ respectively. When event $D$ is created, process P2 is aware of epoch 52 (from P2) and epoch 50 (from P1). It is aware of timestamp $40$ from P3. However, this information is overridden by the clock synchronization guarantee that says that the clock of P3 is at least $47$. Thus, the offsets are set to $[3, e, \epsilon]$
Here, the permissible ordering is $CABD$. 

In this figure, if $\epsilon$ were $20$ then the timestamp of $D$ would be changed to $[3,2,12]$. Furthermore, $B$ and $C$ could be replayed in any order. Thus, the permissible replays would be $CABD$ or $ABCD$ or $ACBD$.

\section{Properties of $\repcl$}
\label{sec:properties}

In this section, first, we define the $<$ relation on two timestamps $\repcl.e$ and $\repcl.f$. Then, we identify the properties of this $<$ relation and the happened-before relation. 

Given timestamps 

$\repcl.e = \br{\maxt.e, \texttt{offset}.e[], \texttt{counter}.e[]}$ and 

$\repcl.f = \br{\maxt.f, \texttt{offset}.f[], \texttt{counter}.f[]}$, 

we say that $\repcl.e < \repcl.f$ iff 

\begin{equation*}
\begin{aligned}
    &\maxt.f > \maxt.e + \clockskew \\
    &\quad \vee \Bigg( |\maxt.f - \maxt.e| \leq \clockskew \\
    &\quad \quad \land \Bigg( \Bigg( \forall l (\maxt.e - \texttt{offset}.e.l) \leq (\maxt.f - \texttt{offset}.f.l) \Bigg) \\
    &\quad \quad \quad \land \Bigg( \exists l (\maxt.e - \texttt{offset}.e.l) < (\maxt.f - \texttt{offset}.f.l) \Bigg) \Bigg) \\
    &\quad \quad \vee \Bigg( \forall l (\maxt.e - \texttt{offset}.e.l) = (\maxt.f - \texttt{offset}.f.l) \\
    &\quad \quad \quad \land \Bigg( \forall l (\texttt{counter}.e.l) \leq (\maxt.f - \texttt{counter}.f.l) \\
    &\quad \quad \quad \quad \land \exists l (\texttt{counter}.e.l) < (\texttt{counter}.f.l) \Bigg) \Bigg) \Bigg)
\end{aligned}
\end{equation*}

The above $<$ relation first compares if $\maxt.f$ and $\maxt.e$ are far apart. If that is the case, we define $\repcl.e < \repcl.f$. If they are close, i.e., $|\maxt.f-\maxt.e|\leq \epsilon$, then, we compare the offsets. Since $\maxt.e-$offset$.e.k$ identifies the knowledge $e$ had about the epoch of process $k$, we use a comparison that is similar to vector clocks to determine if $<$ relation holds between $\repcl.e$ and $\repcl.f$. Finally, if the offsets are also equal then we use the comparison of counters (again in the same fashion as vector clocks).

We overload the $||$ relation for comparing timestamps as well. Specifically, given timestamps $\repcl.e$ and $\repcl.f$, we say that $\repcl.e || \repcl.f$ iff
\begin{equation}
 \neg(\repcl.e < \repcl.f) \wedge \neg(\repcl.f < \repcl.e)        
\end{equation}

From the construction of the timestamp algorithm, we have the following two lemmas:

\paragraph{Lemma 1:}
\label{lemma:repclhappenedbefore}
(e happened before f) $\Rightarrow \repcl.e < \repcl.f$

\paragraph{Lemma 2:} 
\label{lemma:repclconcur}
$|\maxt.e-maxt.f|\leq \clockskew \wedge (e||f)\  \ \ \Rightarrow \ \ \  \repcl.e || \repcl.f$

\paragraph{Requirement 1 of $\repcl$: }
Observe that Lemma 1 satisfies the first requirement of $\repcl$; if $e$ happened before $f$ then $e$ must be replayed before $f$. 

\paragraph{Requirement 2 of $\repcl$: }
Now, we focus on the second requirement. Specifically, we show that by letting $\epsilonone=\clockskew+\intervalsize$, the second requirement is satisfied.

Observe that in the $\repcl$ algorithm, messages carry the epoch values of multiple processes. This allows a process to learn epoch information about other processes. For the subsequent discussion, imagine that the messages also carried the actual physical time as well. In this case, $j$ will learn about the clock of a process $k$ via such messages. Additionally, $j$ will also learn about the clock of a process $k$ based on the assumption of clock synchronization. Likewise, when event $e$ is created, it will have some information about the clock of each process. 
Let $\maxph.e$ and $\maxph.f$ be the maximum clock (of any process) that $e$ and $f$ are aware of when they occurred. 
If $\maxph.f > \maxph.e+\epsilonone$ then $f$ cannot occur before $e$ under the clock synchronization guarantee of $\epsilonone$. Now, we show that in this situation, it is guaranteed that $\repcl.e < \repcl.f$.

By definition of $\maxt$, $\maxt.f = \lfloor \frac{\maxph.f}{\intervalsize} \rfloor$ and  $\maxt.e = \lfloor \frac{mpt.e}{\intervalsize} \rfloor$. Additionally, we have 

\begin{tabbing}
    \hspace*{5mm} \=
    $-1 < (x - \lfloor x \rfloor) - (y - \lfloor y \rfloor) < 1$\\
$\implies$ 
    $-1 < (\frac{\maxph.f}{\intervalsize} - \lfloor \frac{\maxph.f}{\intervalsize}  \rfloor) - (\frac{\maxph.e}{\intervalsize} - \lfloor \frac{\maxph.e}{\intervalsize} \rfloor) < 1$\\
$\implies$
    $-1 < (\frac{\maxph.f}{\intervalsize} - \maxt.f) - (\frac{\maxph.e}{\intervalsize} - \maxt.e) < 1$\\
$\implies$
    $-1 < (\frac{\maxph.f-\maxph.e}{\intervalsize}) - (\maxt.f - \maxt.e)  < 1$\\
$\implies$
    $-\intervalsize \leq ({(\maxph.f-\maxph.e)} - (\maxt.f - \maxt.e)\intervalsize  \leq \intervalsize$\\
\end{tabbing}

Now, if $\maxph.f-\maxph.e > \clockskew+\intervalsize$ then we can rewrite the second inequality as $(\frac{\clockskew+ \intervalsize}{\intervalsize} - (\maxt.f - \maxt.e)  \leq 1$. Using the fact that $\clockskew=\epsilon*\intervalsize$, we have $\epsilon < (\maxt.f-\maxt.e)$. In other words, $\maxph.f-\maxph.e > \clockskew + \intervalsize \Rightarrow (\maxt.f > \maxt.e+\epsilon)$. which gives us $\repcl.e < \repcl.f$. In other words, we have

\paragraph{Lemma 3: } 
\label{lem:effar}
If $f$ occurred far after $e$, i.e., $f$ could not have occurred before $e$ in a system that guarantees that clocks are synchronized within $\epsilonone = \clockskew+\intervalsize$ then $e$ will be replayed before $f$, i.e., $\repcl.e < \repcl.f$.

\paragraph{Requirement 3 of $\repcl$: }
Next, we consider the case where $e$ and $f$ could have occurred in any order if the underlying system guaranteed that clocks were synchronized to be within $\epsilontwo=\clockskew-\intervalsize$. Letting the maximum clock that event $e$ (respectively, $f$) was aware of to be $\maxph.e$ (respectively, $\maxph.f$), we observe that $|\maxph.e - \maxph.f| \leq \epsilontwo$. Furthermore, $e$ and $f$ must be causally concurrent. Under this scenario, we show that $\repcl.e || \repcl.f$.

If $|\maxph.e - \maxph.f| \leq \clockskew-\intervalsize$, we have

\begin{tabbing}
    \hspace*{5mm} \=

$|mpt.e - mpt.f| \leq \clockskew - \intervalsize$\kill

$\implies$
    ${|\frac{mpt.e}{\intervalsize} - \frac{mpt.f}{\intervalsize}| \leq \ \epsilon-1}$ \hspace*{5mm} \= // since $\clockskew=\epsilon*\intervalsize$ \\
$\implies$
    ${|\lfloor \frac{mpt.e}{\intervalsize} \rfloor - \lfloor \frac{mpt.f}{\intervalsize} \rfloor| \leq \ \epsilon}$ \>
since $|(x - \lfloor x \rfloor) - (y - \lfloor y \rfloor)| < 1$\\
$\implies$ 
    $|\maxt.e-\maxt.f|\leq \epsilon$
\> by definition of $\maxt$
\end{tabbing}

Now, from Lemma 2, $\repcl.e || \repcl.f$. In other words, 
\paragraph{Lemma 4: }
\label{lem:allorder}
If $e$ and $f$ could have occurred in any order in a system where clocks were synchronized to be within $\clockskew-\intervalsize$ then $\repcl.e||\repcl.f$.  

\section{Effect of discretization and comparison with Hybrid Vector Clocks \cite{yingchareonthawornchai2018analysis}}
We note that the discretization of the clock via $\intervalsize$ has caused the bounds used for clock synchronization in Lemmas 1 and 2 to be different. We could have eliminated this if we had not discretized the clocks. (Discretization with $\intervalsize$ was not done in \cite{yingchareonthawornchai2018analysis}.) 
However, without discretization, the values of offsets will be very large. Without discretization, we will need to rely on just the physical clocks which have a granularity of under 1 nanosecond. Now, if $\clockskew=1ms$ then the value of the offset could be as large as $10^6$. By discretizing the clock, it would be possible to keep offsets to be very small. We expect that the discretization will not seriously impact the replay. For example, if $\clockskew=1ms$ and $\intervalsize=0.1ms$ then our algorithm will guarantee that causally concurrent events within $0.9ms$ can be replayed in any order. And, events that could not occur simultaneously under a clock synchronization guarantee of $1.1ms$ will be replayed only in one order. Additionally, if $e$ happened before $f$ then $e$ will always be replayed before $f$. 

\section{Representation of the $\repcl$ and its Overhead}
\label{sec:representation}

In this section, we identify how $\repcl$ can be stored to permit efficient computation. As written, $\repcl$ will require $2n+1$ integers. However, a more compact representation will be possible when we account for the fact that it is being used in a system where the clocks are synchronized to be within $\clockskew$. Thus, if $j$ does not hear from $k$ (directly or indirectly) for a long time then the knowledge $j$ would have about the clock of $k$ is the same that is provided by the clock synchronization assumption. In this case, offset$.j.k=\epsilon$. It follows that there is no need to store this value if we interpret \textit{no information about the offset of $k$} to mean that offset$.j.k=\epsilon$.

\begin{figure}
    \centering
    \includegraphics[width=0.9\linewidth]{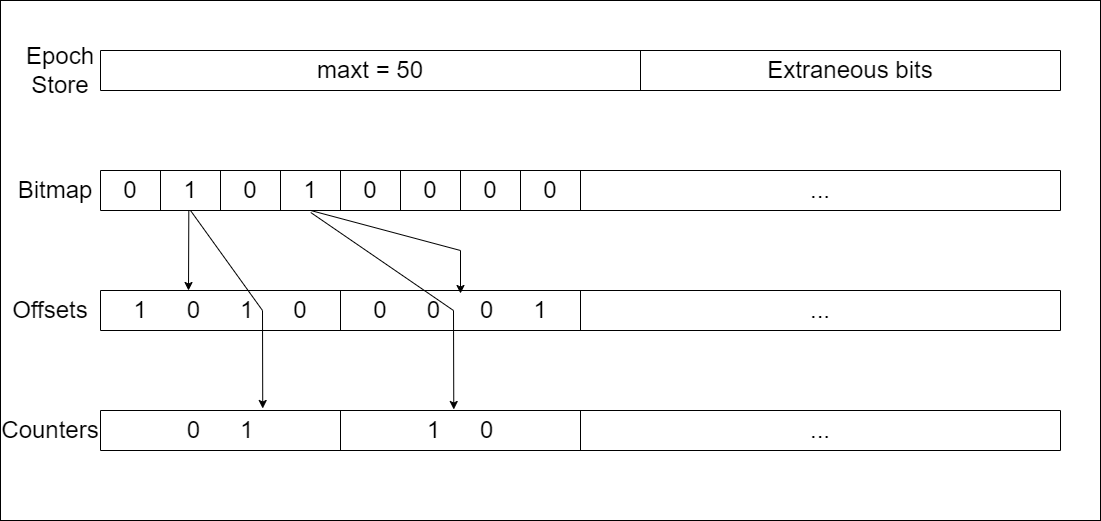}
    \caption{$\repcl$ representation.}
    \label{fig:size_repr}
\end{figure}

With this intuition, we represent $\repcl.j$ as shown in Figure \ref{fig:size_repr}. Here, the value of $\maxt.j$ (represented by the first word) is 50. The next word identifies the bitmap. Since the bit corresponding to process 1 is $0$, it implies that offset$.j.0=\epsilon$. The offset for process 2 is $10$ (first 4 bits of the offset) and $counter.j.2$ is $2$ (first 2 bits of the counter). We note that the bits for each offset and counter are hard-coded based on the system parameters (cf. Chapter \ref{chap:simulation}).

The second word is a bitmap that identifies whether offset$.j.k$ is stored for process $k$. Each offset is stored with a fixed number of bits in the subsequent word(s). 

Next, we show that this representation allows us to reduce the cost of storage as well as the cost of computing timestamps or comparing them (using $<$ relation). Specifically, all these costs are proportional to the number of processes that have communicated with $j$ recently. 
 
With representation in Figure \ref{fig:size_repr}, first, we note that finding the location of the $1$s in the given bitmap can be done in time that is proportional to the number of $1$s in the bitmap. [($(n - (n \& (n - 1))$) will return the number with only the rightmost $1$]. Thus, we have

\paragraph{Observation 1:} 
Shift and MergeSameEpoch can be implemented using $O(x)$ time where $x$ is the number of bits set to $1$ in the bitmap.

Note that this means that the time to compute the timestamp for send/receive at process $j$ is not dependent upon the number of processes in the system. But only processes that have recently communicated with process $j$. In turn, this means that 

\paragraph{Observation 2:} 
Send and Receive can be implemented using $O(x)$ time where $x$ is the number of bits set to $1$ in the bitmap.

\paragraph{Observation 3:} 
Given two timestamps, $\repcl.e$ and $\repcl.f$, we can determine if $\repcl.e < \repcl.f$ in $O(x)$ time where $x$ is the number of bits set to $1$ in $e$ and $f$. 

It follows that the number of bits that are $1$ in a given timestamp identifies not only the storage cost of the timestamps but also the time to compute these timestamps at run time. Effectively, this also identifies the overhead of the timestamps that enable the replay of the computation. Hence, in Chapter \ref{chap:simulation}, we focus on identifying scenarios where the cost of storing these offsets is within the limits identified by the user. 

We note that the above approach will work as long as the number of processes is less than the number of bits in a word (typically, 64 in today's systems). We expect that this will be more than sufficient for many systems in practice. If there are more than 64 processes, we expect that process $j$ is communicating with only a subset of these processes. And, if process $j$ does not communicate with someone there is no need to store offsets for them. Thus, this approach can be extended for the case where the number of processes is larger. However, the details are out of the scope of this thesis. 

\chapter{Simulator Setup}
\label{chap:setup}


In this chapter, we discuss the construction of a custom discrete event simulator, to serve as a validation to state-of-the-art simulators available for research. We first design a baseline simulator build for natively supporting the $\repcl$ infrastructure. We call this the Custom Discrete Event Simulator, or CDES. The goal of the $\repcl$ was to serve as a plug-and-play structure that allows replays to be supported natively by virtue of the clock's algorithm. In other words, the clock should be \textit{self-contained}, where replay should only require $\repcl$-timestamped logs to function correctly, and any simulator should be able to incorporate the visualization just by implementing the clock in its infrastructure. Details of this simulator are further discussed in Section \ref{sec:repsim}.

However, building a simulator for the $\repcl$ introduces a bias in the design. To validate the results to more real-world scenarios, we considered using a state-of-the-art (SOTA) simulator, NS-3 \cite{ns3}. NS-3 proved to be an effective simulator to implement the $\repcl$ structure. However, NS-3 posed challenges in making node-local noisy clocks, which is discussed further in Section \ref{sec:ns3}. For this reason, we revise NS-3 to implement structures that would aid us in approximating what a node-local clock would look like.

Since the NS-3 team expressed the desire to enhance NS-3 with node-local noisy clocks, we decided to build a custom implementation of the same. We designed the node-local clock in a way to permit any arbitrary node level clock (e.g., HLC \cite{hlc}, Vector clock \cite{Fidge87}\cite{mattern1988virtual}, Logical Clocks \cite{Lamport78}, etc.). 

The following chapter is organized as follows. Section \ref{sec:repsim} describes the custom simulator we designed to validate the results obtained by any generic simulator, and identify key differences in the results. During our research, we chose to implement the custom simulator (CDES) first, as the choice of the SOTA simulator was not apparent. Additionally, the SOTA simulator should produce the same results as the CDES, as CDES was built solely for the $\repcl$. The CDES served as a ground truth system, and any architecture we chose would be validated against the results of this simulator. Section \ref{sec:ns3} talks about NS-3, and the revisions that needed to be made to implement our clock infrastructure. We discuss the application implemented, and the complexities that the application had to handle to provide correct results.

\section{\textit{CDES, A Custom Discrete Event Simulator}}
\label{sec:repsim}

In this section, we detail the design of our own custom discrete event simulator (CDES). We modeled processes containing a physical clock $pt$ and the $\repcl$. The design of the CDES is discussed below. 

Each process maintained a vector $msg\_queue$, that queued messages sent to that process. The messages contained the $\repcl$ of the sending process, along with the time the message was to be processed by the receiving process. This receiving time was configured by the sending process by reading the clock of the receiver and adding the message delay to the time. When the receiving process obtained this message, it would compare its physical time $pt$ with the receiving time, and process the message. Each process had a skewing node-local physical clock. We implemented this by randomly advancing the clock of a process based on a seed, and chose not to advance clocks. We maintained the invariant that for no two processes $i$ and $j$, $|pt.i - pt.j| > \clockskew*\intervalsize$, same as the case in NS-3.  

To test the clock, we used the same parameters for $n$, the number of processes, $\clockskew$, $\intervalsize$, $\delta$ and $\alpha$. The message delay is modeled as the average delay experienced in the simulated network, and can vary by some nonzero $\Delta$ in production. In the simulation, at every \textit{clock tick}, a process delivers any messages it is expected to deliver at that clock tick. It also sends a message to other processes based on the message rate $\alpha$. When a message is sent, the corresponding receive event is added to the receiver's queue based on the value of $\delta$. We also compute the actual value of the maximum clock skew observed in the simulation to ensure that if $\clockskew=1ms$ then the worst-case clock skew is indeed $1ms$. The simulation was initialized such that each process started with the same starting clock and at each microsecond time step, each process made a decision to send a message. The process first generated a random number in the range [0, 100], and if this number was lower than $\alpha$, the process elected to send a message to any other process in the simulation or perform a local event.  Each process in the simulation had a uniform chance to send a message with constant delay $\delta$ to any other process in the system. The total messages sent varied with alpha, within the range of [140, 10208] messages over 10,000 steps of the simulation. The parameters we varied are detailed in Table \ref{tab:params}.

\begin{table}[]
\centering
\begin{tabular}{|l|l|l|l|}
\hline
\textbf{Parameter} & \textbf{Minimum Value} & \textbf{Maximum Value} & \textbf{Increments} \\ \hline
$N$          & 32 processes  & 64 processes   & 32 processes    \\ \hline
$\clockskew$ & 10 units      & 1000 units     & 50 units        \\ \hline
$\intervalsize$    & 100 microseconds       & 1000 microseconds      & 50 microseconds     \\ \hline
$\delta$     & 1 microsecond & 8 microseconds & 2x microseconds \\ \hline
$\alpha$     & 10 messages/s & 160 messages/s & 2x messages/s   \\ \hline
\end{tabular}
\caption{\textbf{Parameter configurations for the NS-3 Simulations. We only selected the configurations where $\clockskew*\intervalsize$ \% 1000 == 0 to give us acceptable clock skew limits of {[1ms, 6ms]}.}}
\label{tab:params}
\end{table}


\section{\textit{NS-3 Simulator}}
\label{sec:ns3}


Network Simulator-3 (NS-3) \cite{ns3} provides a generic discrete event simulator, that works on top of different devices implementing applications in different topologies. What makes NS-3 an attractive option to test clock infrastructures is its versatility in types of nodes available, its ease of use in topology configuration and application design, and most importantly, the configurability it offers in designing simulations. The NS-3 infrastructure describes a generic Simulator, that allows different network configurations to be supported and tested by changing a few parameters of this Simulator class. These factors made NS-3 an attractive option to test and incorporate the $\repcl$.

However, NS-3 posed a few challenges. NS-3, as of the time of writing this thesis, does not provide support for node-local noisy clocks. Clocks in NS-3 are synchronized with the top level Simulator, and do not contain their own clock implementations. There have been attempts in creating a node-local noisy clock, but have not been incorporated into the NS-3 infrastructure. Another challenge stems from this. Due to the absence of node-local noisy clocks, NS-3 does not handle clock drifts. Due to the absence of clock drifts, the $\repcl$ would not store any offsets, as all processes would tick in sync each time with the Simulator class. 

Hence, we devised an API to overcome these key challenges. We implemented a node-local noisy clock, which would approximate a node reading its physical time, and added a value $\delta$ to approximate the noise produced by skewing physical clocks. The algorithm for the node-local noisy clock is described in Algorithm \ref{alg:noisyclock}. Here ,$nt.i$ denotes the node-local time of process $i$.

\begin{algorithm}
\caption{Node-Local Noisy Clock: Get Operation}
\label{alg:noisyclock}
\begin{enumerate}
    \item \textbf{Input:} $SimulatorTime$
   \item \quad $nt.i$ = $random(nt.i, SimulatorTime + (\clockskew*\intervalsize))$

    \item \quad \textbf{Return} $nt.i$
\end{enumerate}
\end{algorithm}

As described in Algorithm \ref{alg:noisyclock}, we receive a clock that maintains the relation that for no two processes $i$ and $j$, is $|pt.i - pt.j| > \clockskew*\intervalsize$, but produces clocks that skew with respect to each other. This helps us produce offsets between different processes in a dynamic fashion, and allows us to handle clock drifts.

Using this node-local clock, we design an application in NS-3 called the ReplaySimulatorApplication, with nodes implementing the node-local clocks and the $\repcl$. The $\repcl$ uses the local clock to perform updates on itself. The ReplaySimulatorApplication picks a random node candidate for each nodes and sends a message at intervals defined by the message rate $\alpha$. The channels implemented provide a maximum data rate of 500 Mbps, and a message delay defined by $\delta$. We also provided the option to choose $\clockskew$ and $\intervalsize$ for the purposes of testing the clock sizes. In a more real-world implementation, only the $\intervalsize$ would be changeable by the user. All other parameters would be specified by the distributed system's operating constraints. 

We simulated a distributed environment to test the clock with five parameters - the number of processes ($n$), the maximum allowed clock skew ($\clockskew$), the interval size ($\intervalsize$), the message rate ($\alpha$), and the message delay in microseconds ($\delta$).  We collected results for about 20 seconds for each run. Table \ref{tab:params} defines the variation statistics of each of these parameters.

A key change we made in the NS-3 simulator is the counter storage. We store \textbf{only the sum of counters} in the counter array space. This is explained further in section 7.3.1. The key idea is that there are very few events that actually store counters for all processes, and by storing just the sum of counters of all processes helps us condense the information needed, without the loss of generality in most cases. We do risk missing a few orderings, but it is an acceptable tradeoff as the cases where a lot of counters are stored are rare.

\chapter{Simulation Results}
\label{chap:simulation}

As demonstrated in Section \ref{sec:properties}, the overhead of $\repcl$ depends upon the number of offsets/counters that need to be stored. And, this value depends upon the number of processes that communicate with a given process in the $\clockskew$ time. In other words, the system parameters will determine the size of $\repcl$. In this section, we evaluate the overhead of $\repcl$ via simulation. For the purposes of the results, we denote the offset array size as $\offsetsize$ and the counter array size as $\countersize$.

In the following sections, we outline the effects of changing different parameters both in the custom simulator and in NS-3. Note that the results for the CDES are reported in 32-bit word lengths, and the results for NS-3 are reported in bits, due to the different data collection techniques used for each. 

\section{Effect of Clock Skew ($\clockskew$)}

In this section, we measure the trends in $\clockskew$ while varying the other parameters to see how the $\offsetsize$ and $\countersize$ are affected. We compare each parameter pair-wise with $\clockskew$ to see the effect of the parameter on the clock skew trend with $\offsetsize$.

\subsection{Analysis of Varying Interval Size ($\intervalsize$) with the CDES}

Here, we keep the $\delta$ and $\alpha$ constant to see how $\offsetsize$ and $\countersize$ change with $\clockskew$. 

\begin{figure}[ht]
  \centering

  \begin{subfigure}[b]{0.3\textwidth}
    \includegraphics[width=\linewidth]{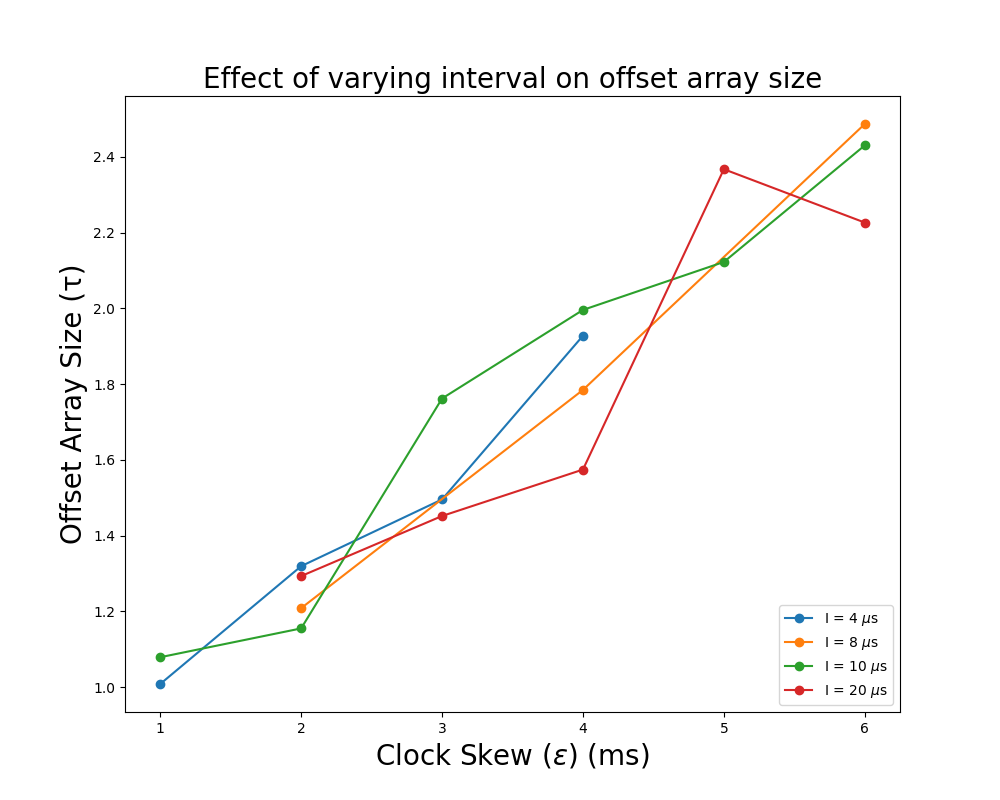}
    \caption{$\alpha$ = 20 msgs/s, $n$ = 32.}
    \label{fig:effectofeps-interval1}
  \end{subfigure}
  \hfill
  \begin{subfigure}[b]{0.3\textwidth}
    \includegraphics[width=\linewidth]{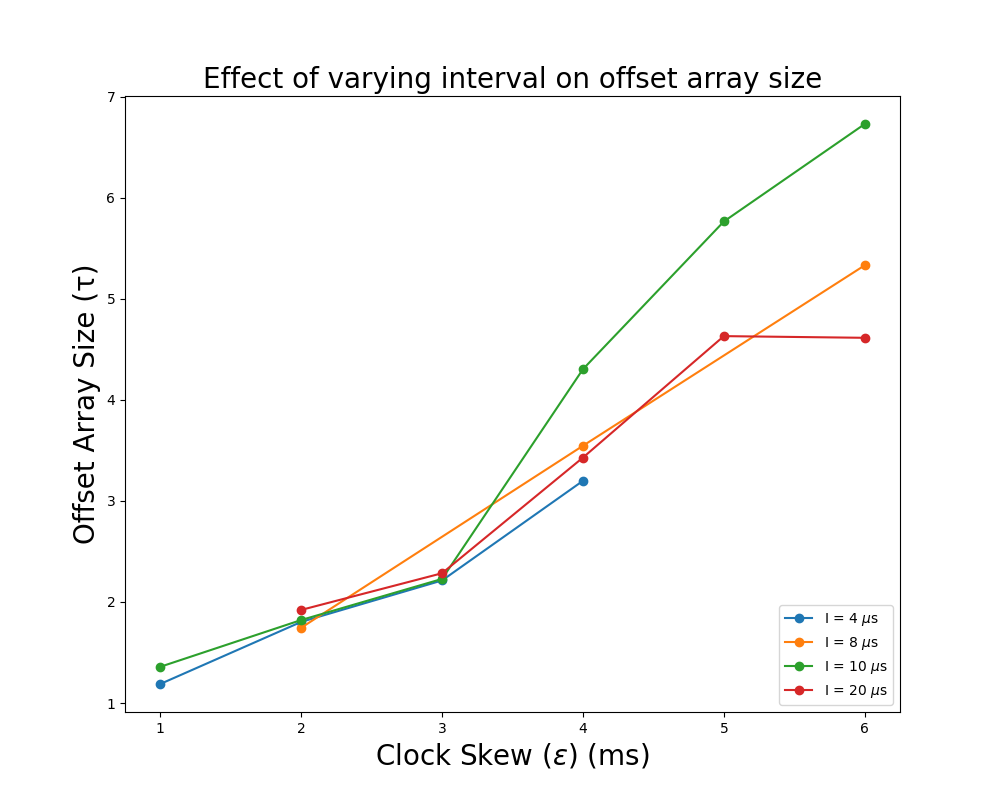}
    \caption{$\alpha$ = 40 msgs/s, $n$ = 32.}
    \label{fig:effectofeps-interval2}
  \end{subfigure}
  \hfill
  \begin{subfigure}[b]{0.3\textwidth}
    \includegraphics[width=\linewidth]{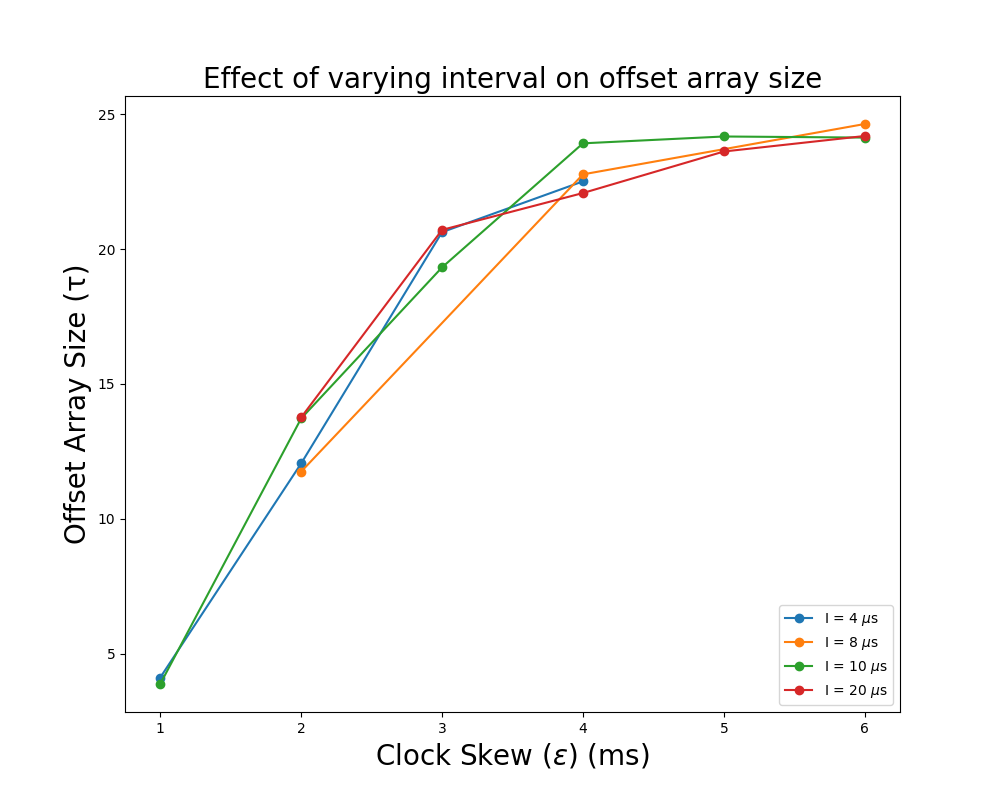}
    \caption{$\alpha$ = 160 msgs/s, $n$ = 32.}
    \label{fig:effectofeps-interval3}
  \end{subfigure}

  \begin{subfigure}[b]{0.3\textwidth}
    \includegraphics[width=\linewidth]{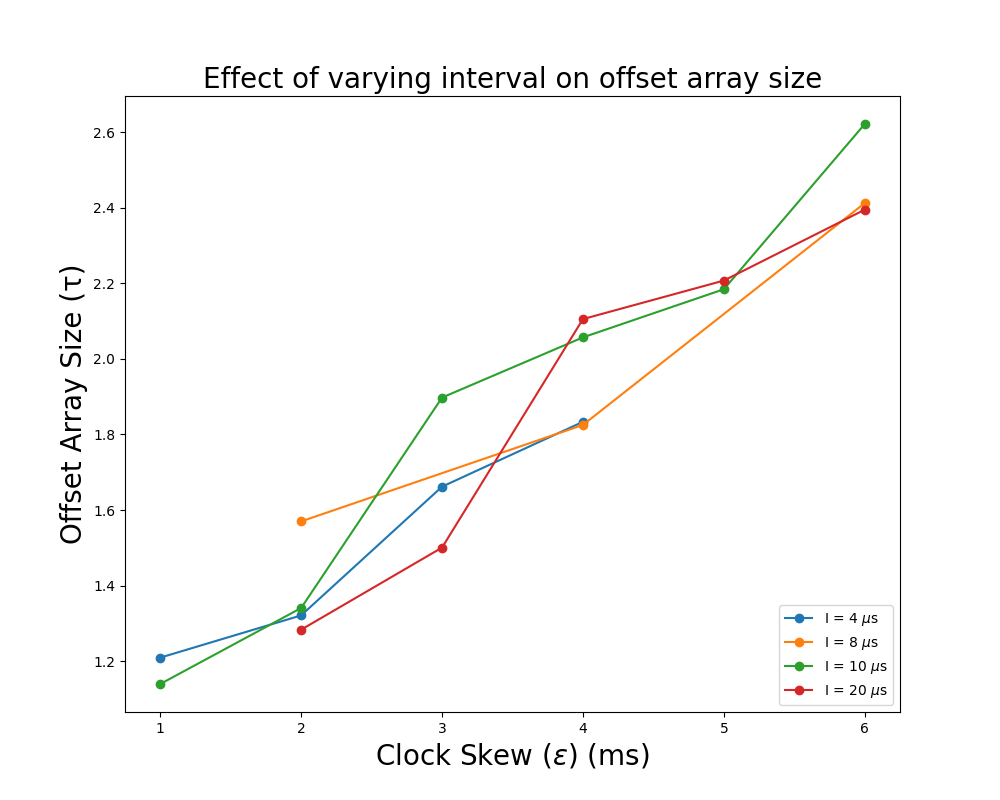}
    \caption{$\alpha$ = 20 msgs/s, $n$ = 64.}
    \label{fig:effectofeps-interval4}
  \end{subfigure}
  \hfill
  \begin{subfigure}[b]{0.3\textwidth}
    \includegraphics[width=\linewidth]{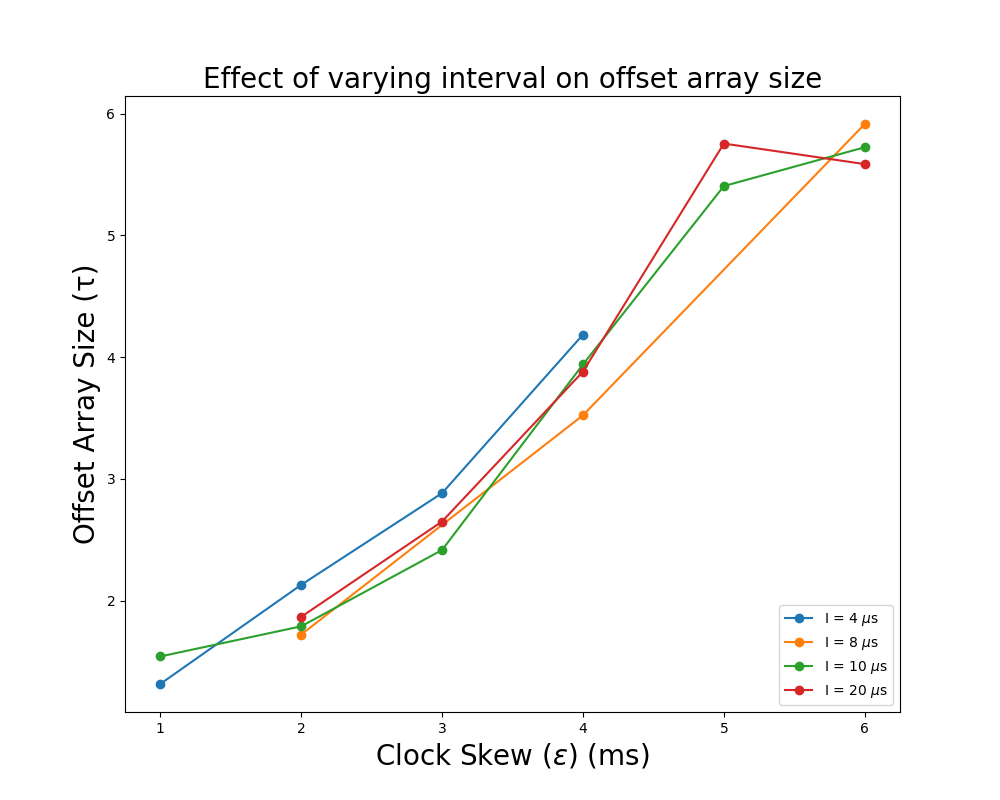}
    \caption{$\alpha$ = 40 msgs/s, $n$= 64.}
    \label{fig:effectofeps-interval5}
  \end{subfigure}
  \hfill
  \begin{subfigure}[b]{0.3\textwidth}
    \includegraphics[width=\linewidth]{diagrams/graphs/IvOff-N32-D8-A160.png}
    \caption{$\alpha$ = 160 msgs/s, $n$ = 64.}
    \label{fig:effectofeps-interval6}
  \end{subfigure}

  \caption{Custom Simulator: $\offsetsize$ vs $\clockskew$ when varying $\intervalsize$, $\delta = 8 \mu s$.}
  \label{fig:effectofeps-intervaloffset}
\end{figure}

\begin{figure}[ht]
  \centering

  \begin{subfigure}{0.4\textwidth}
    \includegraphics[width=\linewidth]{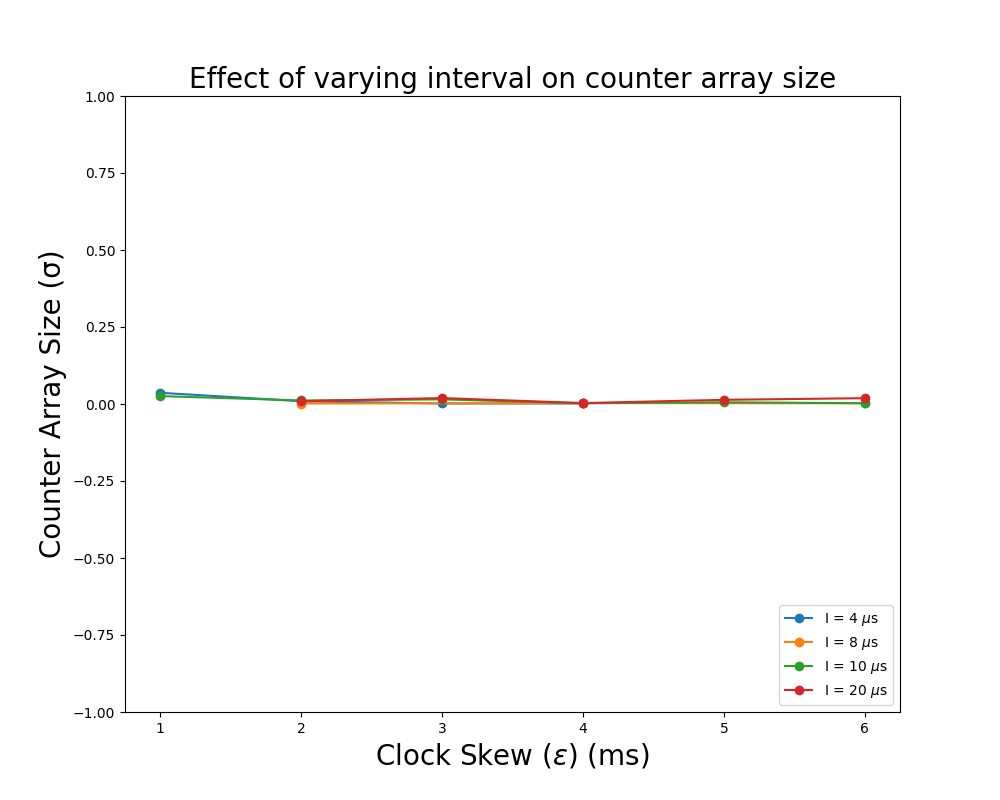}
    \caption{$n$ = 32.}
    \label{fig:effectofeps-interval7}
  \end{subfigure}
  \begin{subfigure}{0.4\textwidth}
    \includegraphics[width=\linewidth]{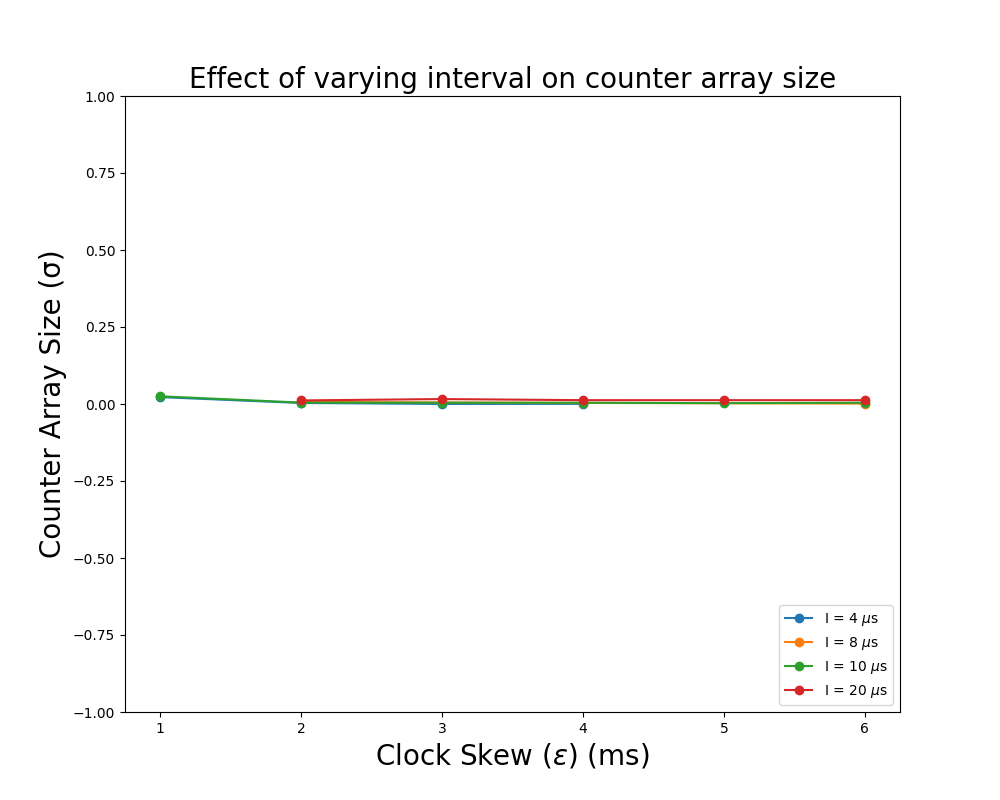}
    \caption{$n$ = 64.}
    \label{fig:effectofeps-interval8}
  \end{subfigure}

  \caption{Custom Simulator: $\countersize$ vs $\clockskew$ when varying $\intervalsize$, $\delta$ = 8 $\mu s$, $\alpha$ = 160 msgs/s.}
  \label{fig:effectofeps-intervalcounter}
\end{figure}

\begin{itemize}
    \item In case of $\offsetsize$ vs $\clockskew$ curve, we notice that the value of $\intervalsize$ has little bearing on $\offsetsize$. As expected, as the value of $\clockskew$ increases, $\offsetsize$ increases with it. This is true for all values of $\alpha$, and we consistently store more offsets as $\alpha$ increases. For any given value of $\alpha$, however, the value of $\intervalsize$ can be chosen to set the granularity of the user's choice and would allow more flexibility in the clock information. Regardless of the choice of $\intervalsize$ by the user, the offset sizes increase with roughly the same trend. These trends are illustrated in Figure \ref{fig:effectofeps-intervaloffset}. 
    \item In the case of $\countersize$ vs $\clockskew$ curve, we see not too much of a variation in $\countersize$, as most events reach a different epoch. On average, we do not see many events storing counters; roughly $0.78\%$ of events store counters, and the values of such counters do not exceed 5 in most cases. Hence, we would need very little space to store these counters. These trends are illustrated in Figure \ref{fig:effectofeps-intervalcounter}. Since this observation is true for all simulations in this paper, we do not discuss the analysis for $\countersize$ in the subsequent sections.
\end{itemize}

\subsection{Analysis of Varying Interval Size ($\intervalsize$) with NS-3}

\begin{figure}[ht]
  \centering

  \begin{subfigure}[b]{0.3\textwidth}
    \includegraphics[width=\linewidth]{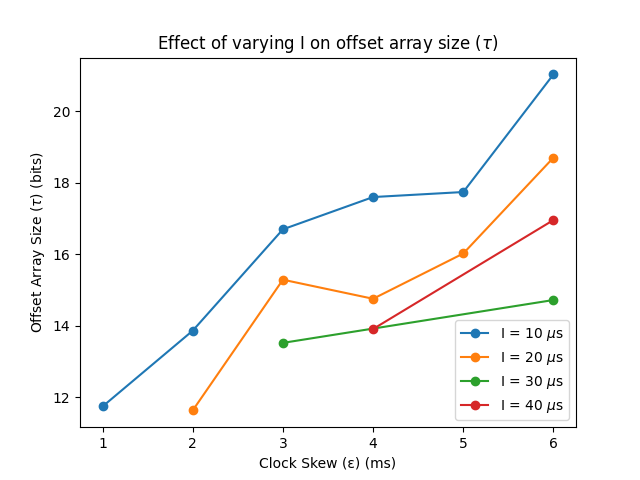}
    \caption{$\alpha$ = 20 msgs/s, $n$ = 32.}
    \label{fig:ns3-effectofeps-interval1}
  \end{subfigure}
  \hfill
  \begin{subfigure}[b]{0.3\textwidth}
    \includegraphics[width=\linewidth]{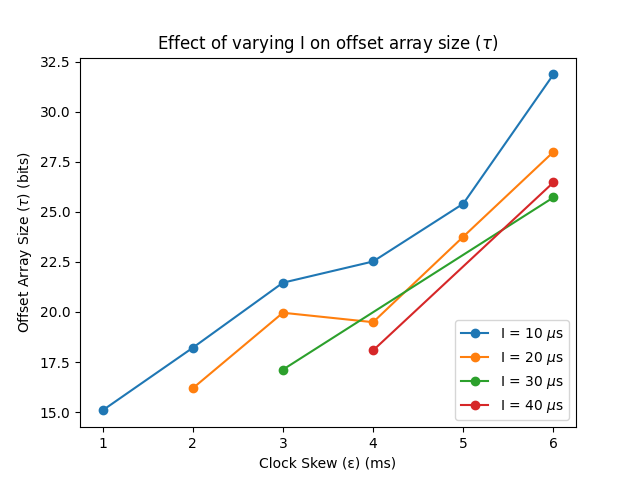}
    \caption{$\alpha$ = 40 msgs/s, $n$ = 32.}
    \label{fig:ns3-effectofeps-interval2}
  \end{subfigure}
  \hfill
  \begin{subfigure}[b]{0.3\textwidth}
    \includegraphics[width=\linewidth]{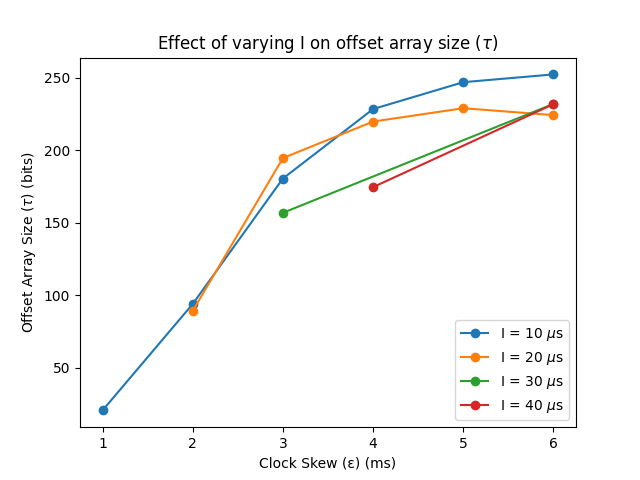}
    \caption{$\alpha$ = 160 msgs/s, $n$ = 32.}
    \label{fig:ns3-effectofeps-interval3}
  \end{subfigure}

  \begin{subfigure}[b]{0.3\textwidth}
    \includegraphics[width=\linewidth]{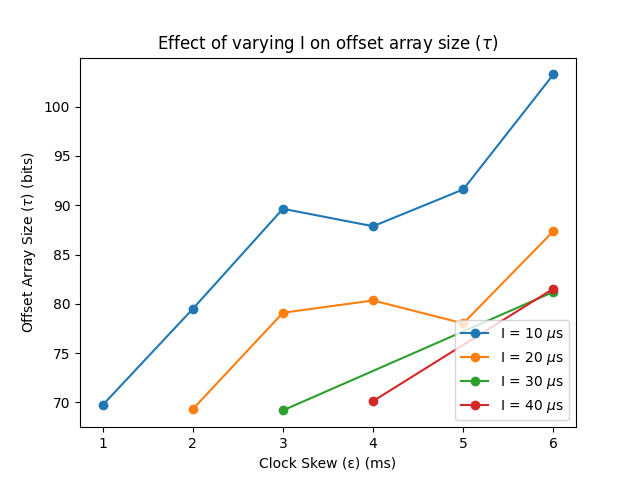}
    \caption{$\alpha$ = 20 msgs/s, $n$ = 64.}
    \label{fig:ns3-effectofeps-interval4}
  \end{subfigure}
  \hfill
  \begin{subfigure}[b]{0.3\textwidth}
    \includegraphics[width=\linewidth]{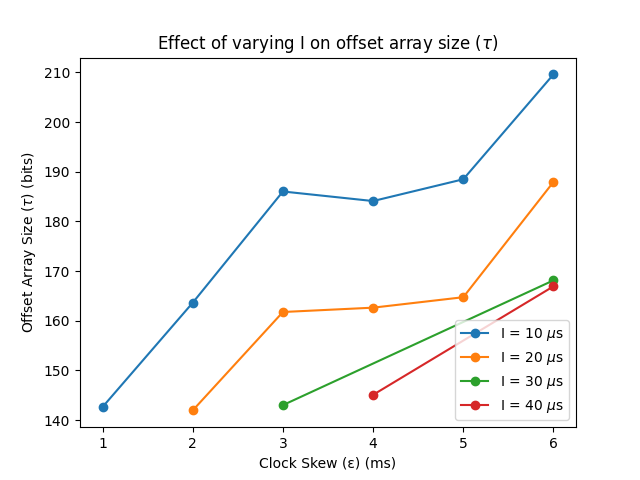}
    \caption{$\alpha$ = 40 msgs/s, $n$= 64.}
    \label{fig:ns3-effectofeps-interval5}
  \end{subfigure}
  \hfill
  \begin{subfigure}[b]{0.3\textwidth}
    \includegraphics[width=\linewidth]{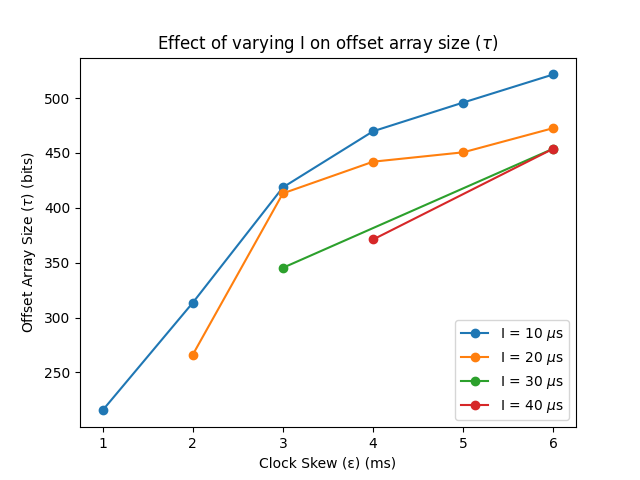}
    \caption{$\alpha$ = 160 msgs/s, $n$ = 64.}
    \label{fig:ns3-effectofeps-interval6}
  \end{subfigure}

  \caption{NS-3 Simulator: $\offsetsize$ vs $\clockskew$ when varying $\intervalsize$, $\delta = 1 \mu s$.}
  \label{fig:ns3-effectofeps-intervaloffset}
\end{figure}

\begin{figure}[ht]
  \centering

  \begin{subfigure}{0.4\textwidth}
    \includegraphics[width=\linewidth]{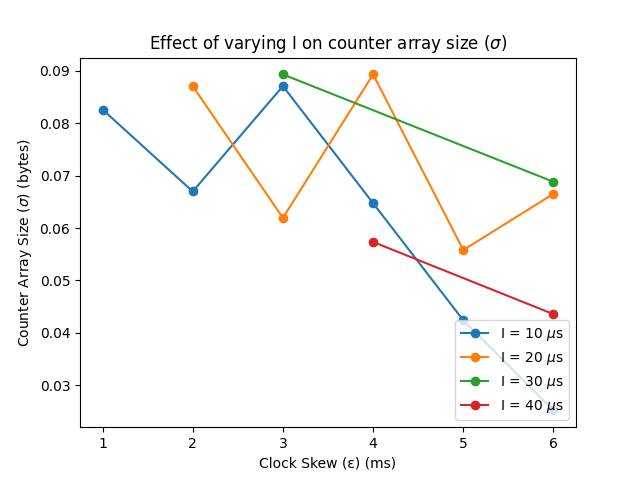}
    \caption{$n$ = 32.}
    \label{fig:ns3-effectofeps-interval7}
  \end{subfigure}
  \begin{subfigure}{0.4\textwidth}
    \includegraphics[width=\linewidth]{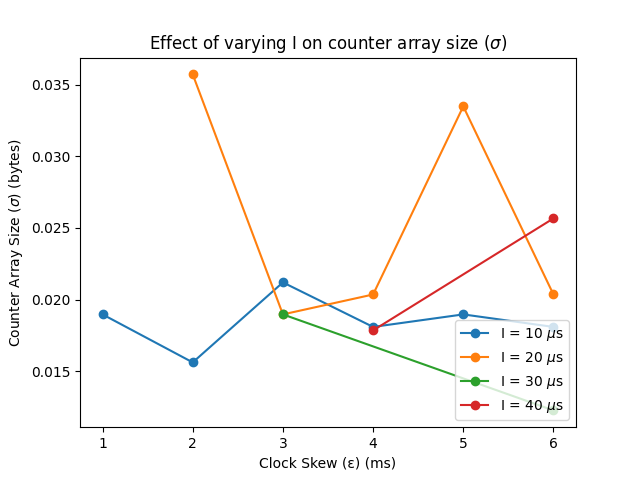}
    \caption{$n$ = 64.}
    \label{fig:ns3-effectofeps-interval8}
  \end{subfigure}

  \caption{NS-3 Simulator: $\countersize$ vs $\clockskew$ when varying $\intervalsize$, $\delta$ = 1 $\mu s$, $\alpha$ = 20 msgs/s.}
  \label{fig:ns3-effectofeps-intervalcounter}
\end{figure}

\begin{itemize}
    \item In the case of the $\offsetsize$ vs $\clockskew$ curve in the NS-3 simulation, we observe a similar trend of $\offsetsize$ increasing with $\clockskew$. This is in agreement with our findings from the custom simulator, and true for all values of $\alpha$. These trends are illustrated in Figure \ref{fig:ns3-effectofeps-intervaloffset}. While we see a decrease in number of bits stored as $\intervalsize$ decreases, the difference is only significant in some cases, notably in the case of lower message rates. At higher message rates, the gap reduces. This is due to the amount of communication happening, and most processes tend to store close to the acceptable limit of the number of offsets we want to store. 
    \item In the case of the $\countersize$ vs $\clockskew$ curve in the NS-3 simulation, we store only the sum of counters. Hence, we see some variations in counter size. With the increase of epsilon, we store slightly higher counters, and this does not change with variation in $\intervalsize$. However, the sizes of the counters vary only from 0.03 bytes in the lowest case of $n$ = 32 to 0.1 bytes in the highest case. Hence, the number stored as the counter value is not too large. This is true even for the case of $n$ = 64, and is depicted in Figure \ref{fig:ns3-effectofeps-intervalcounter}.
\end{itemize}

Since the total size of $\repcl$ depends upon the number of bits for each offset and the total number of offsets, we consider a specific example here. For Figure \ref{fig:size_repr}, the number of offsets is 2 and the size of each offset is 4 bits. Therefore, one word is sufficient to store offsets. Likewise, one word is enough for counters. Thus, we need a total of 4 words to store this timestamp.

(Note that the counters can be stored in the same amount of memory as we have a number of extraneous bits in this representation, specifically in the max epoch word if we elect to store the sum of all these counters here. By doing this, we would lose some information, but considering that the number of events that record meaningful counters is low, this may be an acceptable trade-off.) It is straightforward to observe that the size of $\repcl$ grows linearly with the number of offsets in it. And, the total size of $\repcl$ will require the use of the floor function to identify the number of words necessary to store it. Since the floor operation loses some of the relevant data, we present the value of $\offsetsize$ in this section. 

\subsection{Analysis of Varying Message Delay ($\delta$) for the CDES} 

Here,  we fix the $\intervalsize$ and $\alpha$, and for different $\delta$ values, and we identify how $\offsetsize$ changes with $\clockskew$.

\begin{figure}[ht]
  \centering

  \begin{subfigure}{0.4\textwidth}
    \includegraphics[width=\linewidth]{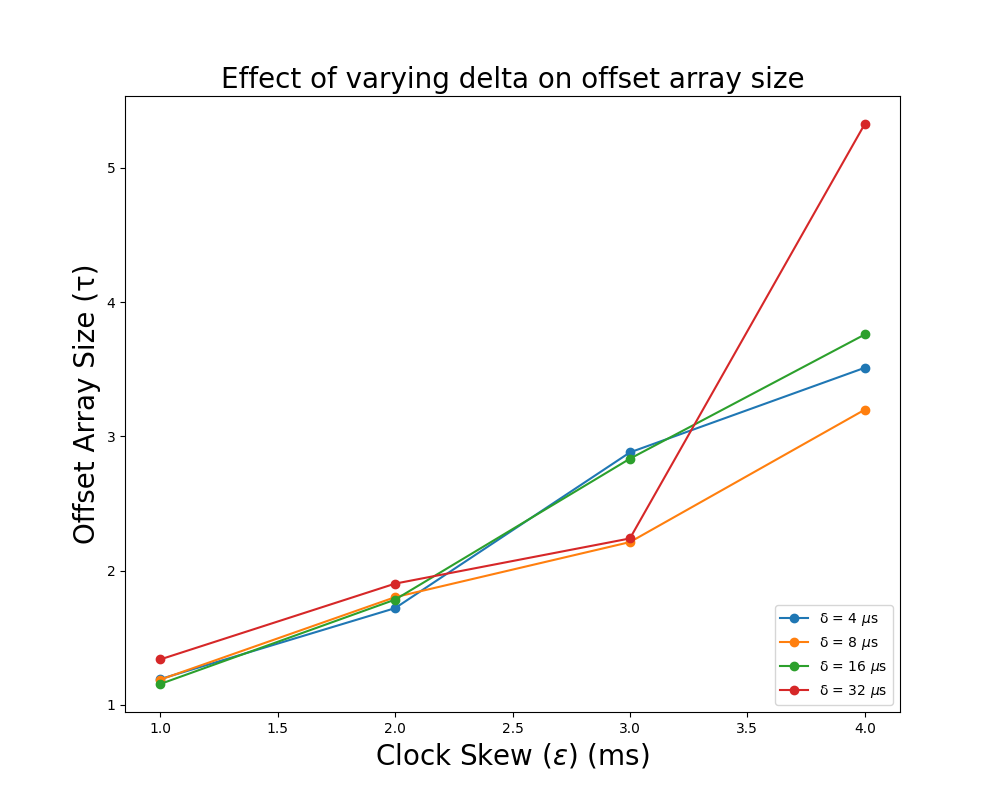}
    \caption{$n$ = 32.}
    \label{fig:effectofeps-delta1}
  \end{subfigure}
  \hfill
  \begin{subfigure}{0.4\textwidth}
    \includegraphics[width=\linewidth]{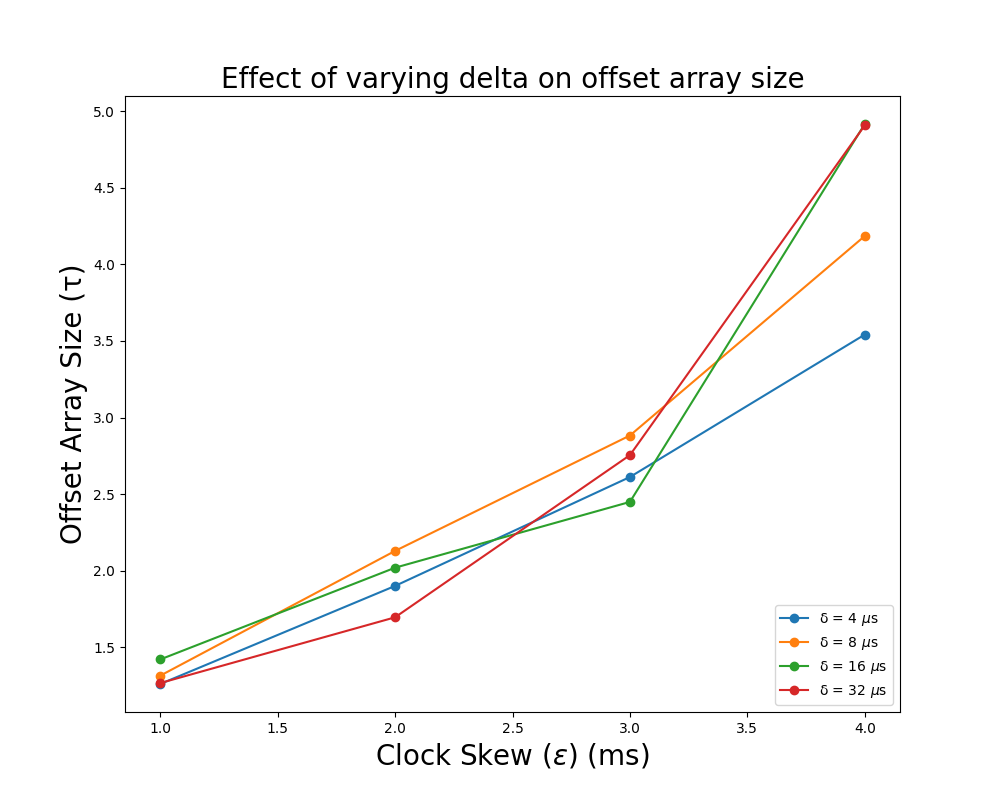}
    \caption{$n$ = 64.}
    \label{fig:effectofeps-delta2}
  \end{subfigure}

  \caption{Custom Simulator: $\offsetsize$ vs $\clockskew$ when varying $\delta$, $\intervalsize$ = 8 $\mu s$, $\alpha$ = 40 msgs/second.}
  \label{fig:effectofeps-deltaoff}
\end{figure}

As $\clockskew$ increases, we see higher values of $\offsetsize$, implying a higher number of offsets stored on average. We observe that higher values of $\delta$ produce a lower number of offsets in each case, barring some noise. This is expected as an increase in $\delta$ implies messages would reach in a delayed fashion, and would lead to processes setting other process offsets to $\epsilon$ due to non-receipt of messages.
As the $\clockskew$ increases, a process hears from more processes (directly or indirectly) within time $\clockskew$. Hence, the number of offsets increases.
This is illustrated in Figure \ref{fig:effectofeps-deltaoff}. 

\subsection{Analysis of Varying Message Delay ($\delta$) for the NS-3} 

\begin{figure}[ht]
  \centering

  \begin{subfigure}{0.4\textwidth}
    \includegraphics[width=\linewidth]{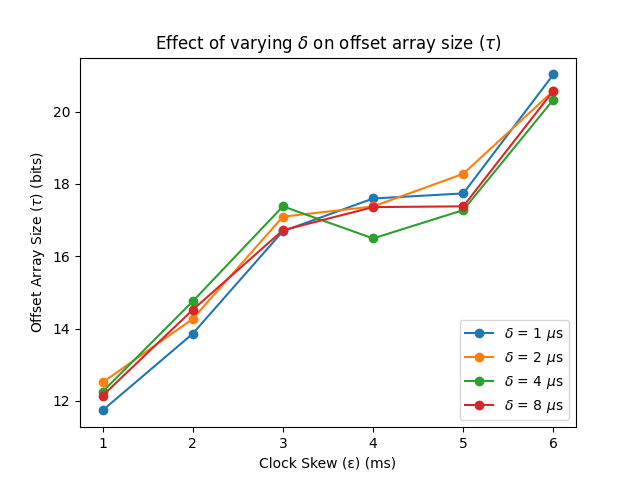}
    \caption{$n$ = 32.}
    \label{fig:ns3-effectofeps-delta1}
  \end{subfigure}
  \hfill
  \begin{subfigure}{0.4\textwidth}
    \includegraphics[width=\linewidth]{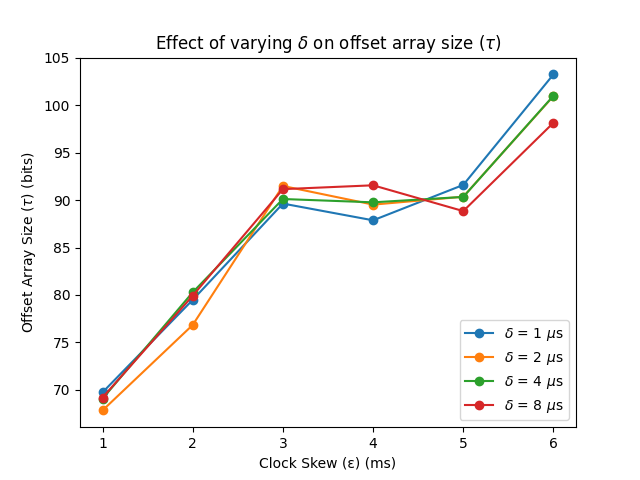}
    \caption{$n$ = 64.}
    \label{fig:ns3-effectofeps-delta2}
  \end{subfigure}

  \caption{NS-3 Simulator: $\offsetsize$ vs $\clockskew$ when varying $\delta$, $\intervalsize$ = 20 $\mu s$, $\alpha$ = 160 msgs/second.}
  \label{fig:ns3-effectofeps-deltaoff}
\end{figure}

As in the case of the Custom simulator, we see higher values of $\offsetsize$ as the $\clockskew$ increases. We do not see too much of a difference as $\delta$ varies however, where the number of bits stored for each $\delta$ value are between $\pm$ 1 bit. The increase in offset size is somewhat linear for the most part as the $\clockskew$ increases, which is the same as observed in the analysis of varying $\intervalsize$. The $\delta$ variations are not pronounced as much due to the fact that the clocks implicitly synchronize when messages are sent between each other. A message sent from far back into the past effectively does not change the clock of the receiver. If a sender gets a clock from the future (in its local observation), it modifies its own clock to push forward to this future timestamp to guarantee the acceptable $\clockskew$ limit. This allows different $\delta$ values to show close to no variation. This is illustrated in Figure \ref{fig:ns3-effectofeps-deltaoff}. It is important to note, however, that when $\delta$ exceeds the $\clockskew$ limit, no process stores any offsets.

\subsection{Analysis of Varying Message Rate ($\alpha$) for the CDES}

Here, we fix the $\intervalsize$ and $\delta$, and for different $\alpha$ values, we identify how $\offsetsize$  changes with $\clockskew$.

\begin{figure}[ht]
  \centering

  \begin{subfigure}{0.4\textwidth}
    \includegraphics[width=\linewidth]{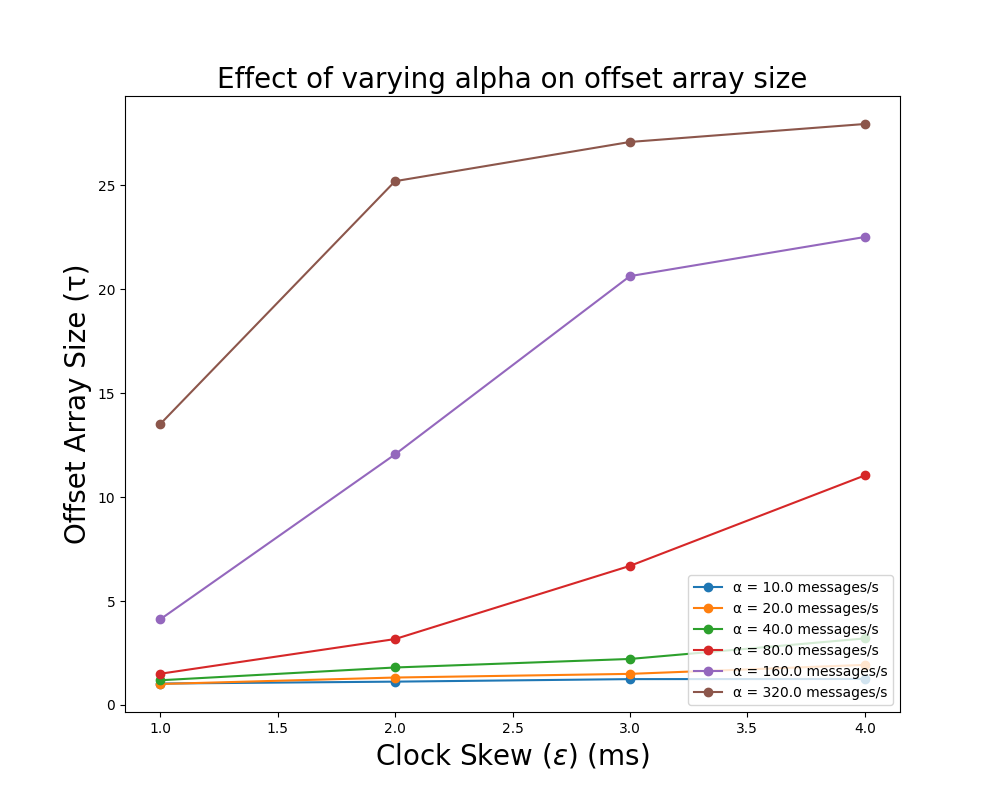}
    \caption{$n$ = 32.}
    \label{fig:effectofeps-alpha1}
  \end{subfigure}
  \hfill
  \begin{subfigure}{0.4\textwidth}
    \includegraphics[width=\linewidth]{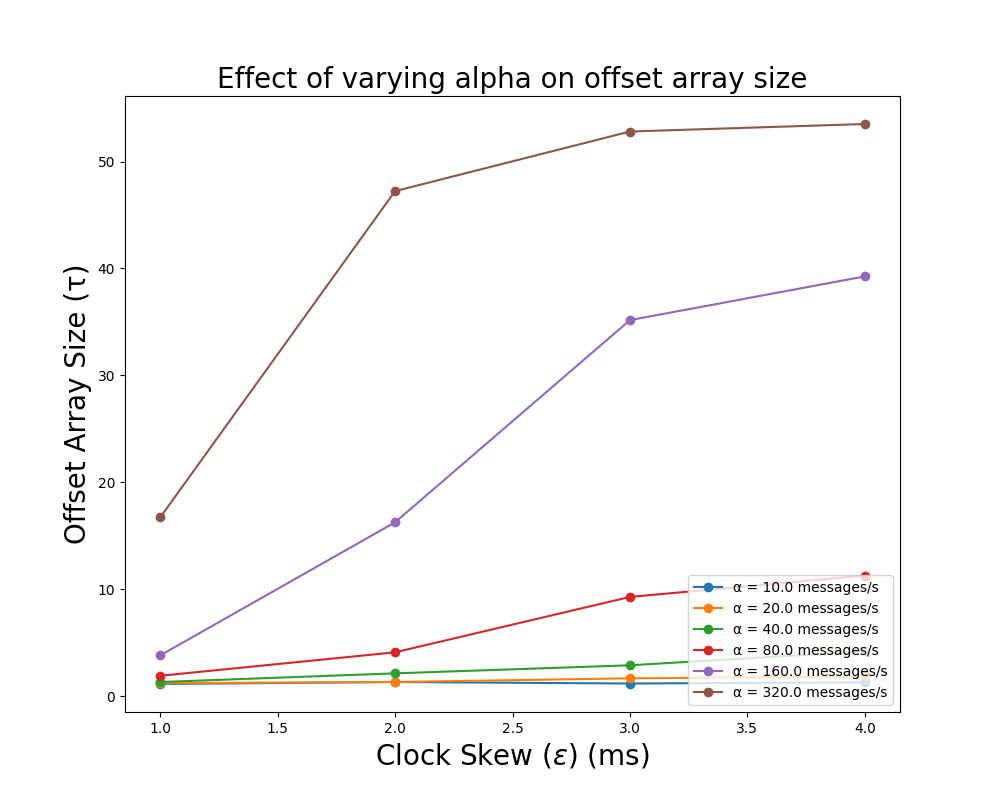}
    \caption{$n$ = 64.}
    \label{fig:effectofeps-alpha2}
  \end{subfigure}

  \caption{Custom Simulator: $\offsetsize$ vs $\clockskew$ when varying $\alpha$, $\intervalsize$ = 4 $\mu s$, $\delta$ = 8 $\mu s$.}
  \label{fig:effectofeps-alphaoff}
\end{figure}

As expected, for lower values of $\alpha$, we consistently store lower offsets, as communication between processes is sporadic. As the $\clockskew$ increases, $\offsetsize$ increases linearly until the bound of $n$ is reached. This is due to the same reason mentioned earlier, as the bound lengths on epochs are larger, even sporadic messages tend to store more offsets on other processes, causing the overall value of $\offsetsize$ to increase. This is illustrated in Figure \ref{fig:effectofeps-alphaoff}.

\subsection{Analysis of Varying Message Rate ($\alpha$) for NS-3}

\begin{figure}[ht]
  \centering

  \begin{subfigure}{0.4\textwidth}
    \includegraphics[width=\linewidth]{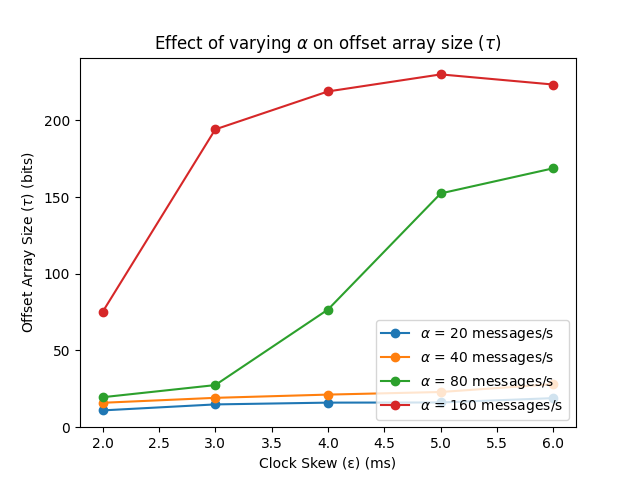}
    \caption{$n$ = 32.}
    \label{fig:ns3-effectofeps-alpha1}
  \end{subfigure}
  \hfill
  \begin{subfigure}{0.4\textwidth}
    \includegraphics[width=\linewidth]{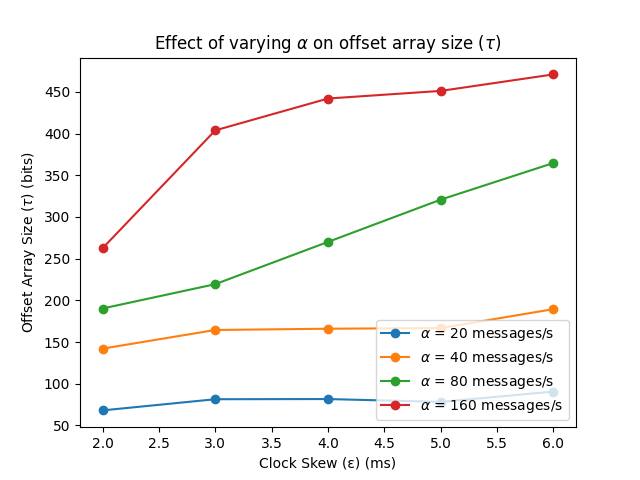}
    \caption{$n$ = 64.}
    \label{fig:ns3-effectofeps-alpha2}
  \end{subfigure}

  \caption{NS-3 Simulator: $\offsetsize$ vs $\clockskew$ when varying $\alpha$, $\intervalsize$ = 20 $\mu s$, $\delta$ = 4 $ms$.}
  \label{fig:ns3-effectofeps-alphaoff}
\end{figure}

Our results from the custom simulator are confirmed by the experiments in NS-3, where lower values of $\alpha$ store lower offsets due to lower communication. In the case of NS-3, higher values of $\alpha$ store many more offsets than what we would like our upper limit to be (about one word of offsets stored per clock). This is guaranteed by having lower values of $\alpha$. This is illustrated in Figure \ref{fig:ns3-effectofeps-alphaoff}.

\section{Effect of Interval Size($\intervalsize$)}

In this section, we observe the trends in $\intervalsize$ with respect to $\delta$ and $\alpha$. 

\subsection{Analysis of Varying Message Delay ($\delta$) for the CDES}

Here, we fix the $\clockskew$ and $\alpha$ and check how $\offsetsize$ changes with $\intervalsize$. 

\begin{figure}[ht]
  \centering

  \begin{subfigure}{0.4\textwidth}
    \includegraphics[width=\linewidth]{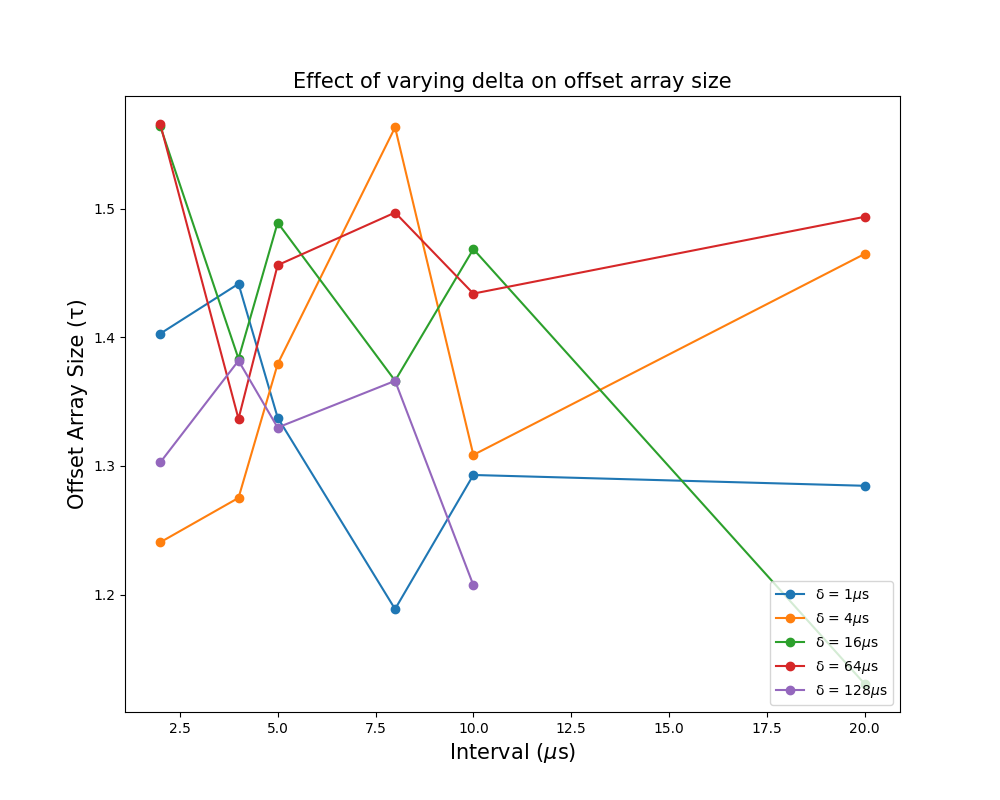}
    \caption{$n$ = 32.}
    \label{fig:effectofint-delta1}
  \end{subfigure}
  \hfill
  \begin{subfigure}{0.4\textwidth}
    \includegraphics[width=\linewidth]{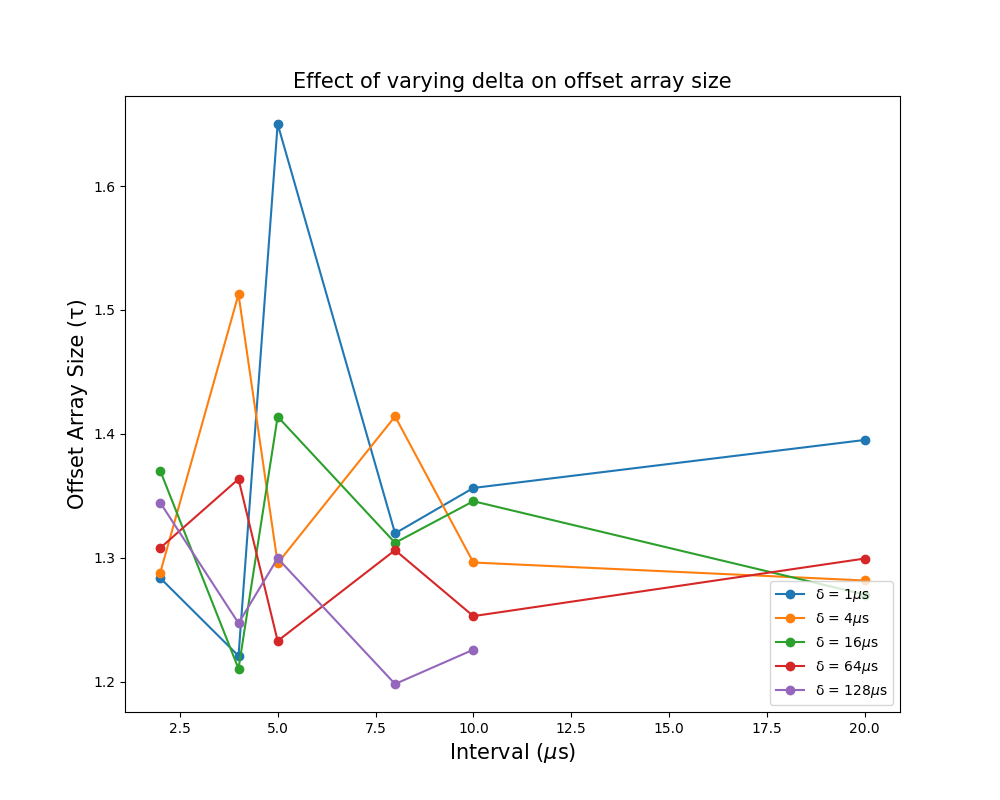}
    \caption{$n$ = 64.}
    \label{fig:effectofint-delta2}
  \end{subfigure}

  \caption{Custom Simulator: $\offsetsize$ vs $\intervalsize$ when varying $\delta$, $\clockskew$ = 2 ms, $\alpha$ = 20 msgs/s.}
  \label{fig:effectofint-deltaoff}
\end{figure}

From Figure \ref{fig:effectofint-deltaoff}, we observe that the value of $\intervalsize$ does not really have a significant effect on $\offsetsize$ (Note that the $Y$ axis of this figure varies only from $1.2$ to $1.5$.) This means that the selection of $\intervalsize$ does not affect the number of offsets maintained by a process. However, it affects the size of each offset. Specifically, the max value of the offset is $\epsilon=\frac{\clockskew}{\intervalsize}$ and the number of bits required for each offset is $\log_2 \epsilon$. Hence, a larger value of $\intervalsize$ is better for reducing the size of the $\repcl$. However, with larger $\intervalsize$, the guarantees provided by $\repcl$ are lower. Specifically, Lemma 3 shows that some unforced reordering may occur when events $e$ and $f$ differ by time $\clockskew+\intervalsize$. Users should therefore choose the value of $\intervalsize$ based on the desired guarantees of $\repcl$ or the maximum desirable offset.

\subsection{Analysis of Varying Message Delay ($\delta$) for NS-3}

In the case of the NS-3 simulator, we observed that the size of offsets decreased linearly with increase in $\intervalsize$. This is attributed to more information being stored in counters, as the length of the interval increases, and less offsets being stored. The likelihood that all processes are in the same epoch increases as the $\intervalsize$ increases, leading to lower number of offsets stored. The variation with $\delta$ is negligible, and seemingly random, which is confirmed with the custom simulator results.

\begin{figure}[ht]
  \centering

  \begin{subfigure}{0.4\textwidth}
    \includegraphics[width=\linewidth]{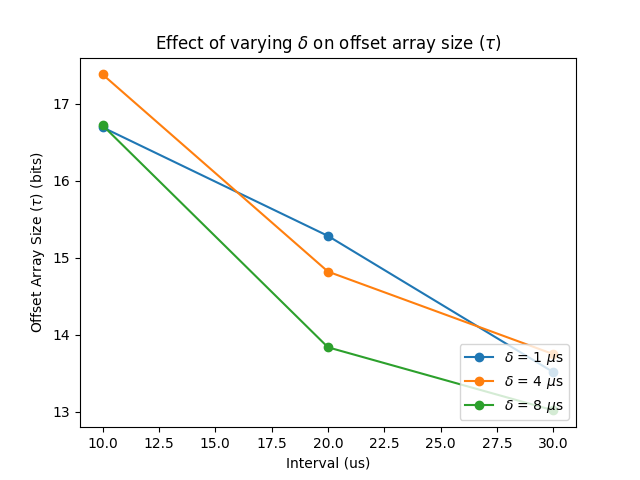}
    \caption{$n$ = 32.}
    \label{fig:ns3-effectofint-delta1}
  \end{subfigure}
  \hfill
  \begin{subfigure}{0.4\textwidth}
    \includegraphics[width=\linewidth]{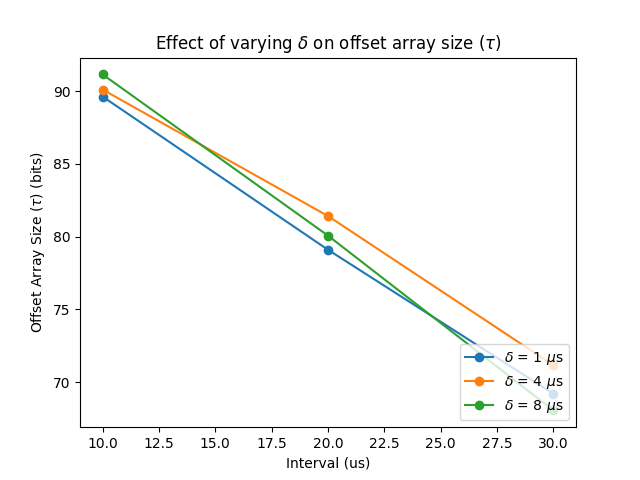}
    \caption{$n$ = 64.}
    \label{fig:ns3-effectofint-delta2}
  \end{subfigure}

  \caption{NS-3 Simulator: $\offsetsize$ vs $\intervalsize$ when varying $\delta$, $\clockskew$ = 3 ms, $\alpha$ = 20 msgs/s.}
  \label{fig:ns3-effectofint-deltaoff}
\end{figure}

\subsection{Analysis of Varying Message Rate ($\alpha$) for the CDES}

\begin{figure}[ht]
  \centering
  \begin{subfigure}{0.4\textwidth}
    \includegraphics[width=\linewidth]{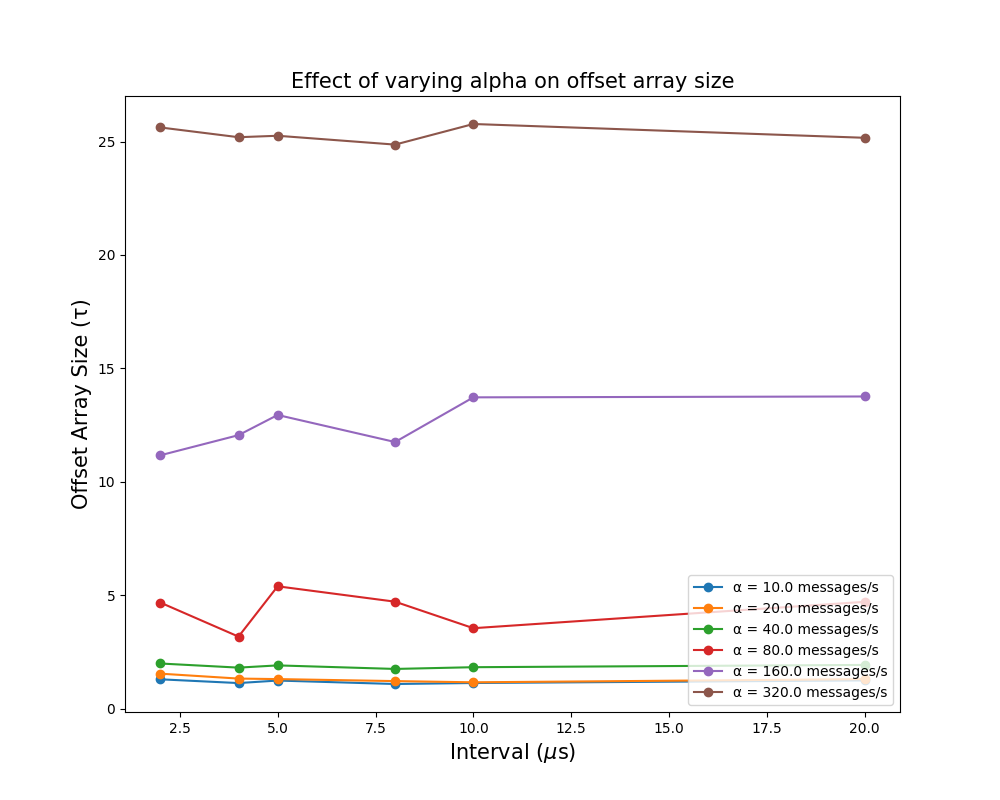}
    \caption{$n$ = 32.}
    \label{fig:effectofint-alpha1}
  \end{subfigure}
  \hfill
  \begin{subfigure}{0.4\textwidth}
    \includegraphics[width=\linewidth]{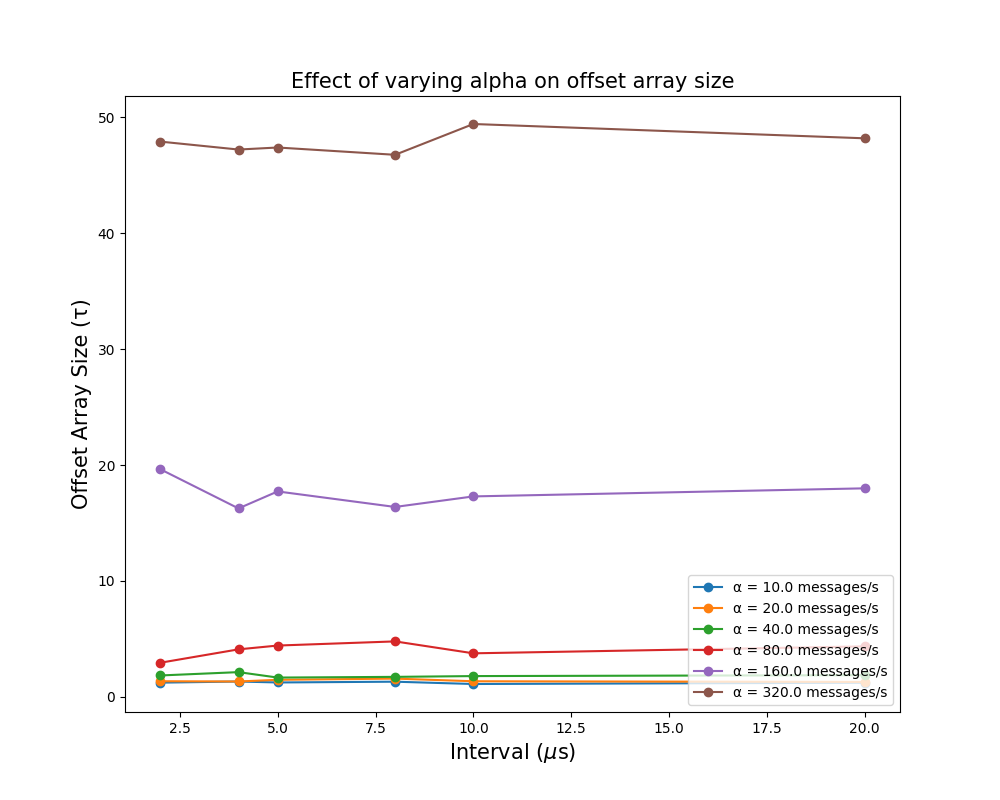}
    \caption{$n$ = 64.}
    \label{fig:effectofint-alpha2}
  \end{subfigure}
  \caption{Custom Simulator: $\offsetsize$ vs $\intervalsize$ when varying $\alpha$, $\clockskew$ = 2 ms, $\delta$ = 8 $\mu s$.}
  \label{fig:effectofint-alphaoff}
\end{figure}

For every point in the $\intervalsize-\offsetsize$ trend, lower $\alpha$ values produce lower $\offsetsize$. This is consistent with our observations so far as communication is infrequent and processes tend to not hear from other processes, subsequently not storing their offsets. As $\intervalsize$ increases, the $\offsetsize$ remains the same, as the $\delta$ remains the same. When $\delta$ is constant, and  $\clockskew$ is constant, messages being sent either remain in the same epoch ($\maxt$) as they would in the previous $\intervalsize$ chosen, or a message that was between different intervals in a lower $\intervalsize$, now would be in the same interval. Overall, this does not change the total $\offsetsize$. This is illustrated in Figure \ref{fig:effectofint-alphaoff}.

\subsection{Analysis of Varying Message Rate ($\alpha$) for NS-3}

\begin{figure}[ht]
  \centering
  \begin{subfigure}{0.4\textwidth}
    \includegraphics[width=\linewidth]{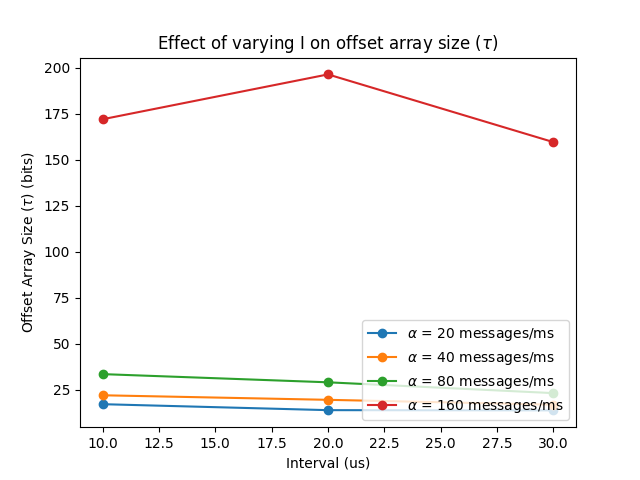}
    \caption{$n$ = 32.}
    \label{fig:ns3-effectofint-alpha1}
  \end{subfigure}
  \hfill
  \begin{subfigure}{0.4\textwidth}
    \includegraphics[width=\linewidth]{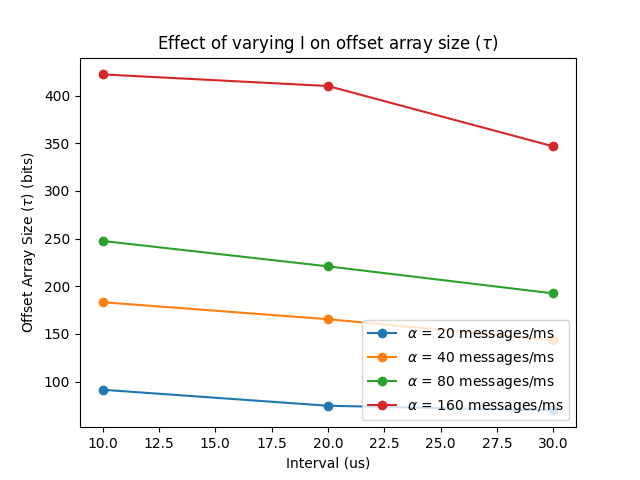}
    \caption{$n$ = 64.}
    \label{fig:ns3-effectofint-alpha2}
  \end{subfigure}
  \caption{NS-3 Simulator: $\offsetsize$ vs $\intervalsize$ when varying $\alpha$, $\clockskew$ = 3 $ms$, $\delta$ = 2 $ms$.}
  \label{fig:ns3-effectofint-alphaoff}
\end{figure}

As observed in the custom simulator, lower $\alpha$ values produce lower $\offsetsize$, due to infrequency in communication. We additionally observe that the $\offsetsize$ remains constant with increasing $\intervalsize$, consistent with our results from the CDES simulator in the previous section. This is illustrated in Figure \ref{fig:ns3-effectofint-alphaoff}.

\section{Effect of Message Delay($\delta$)}

In this section, we observe the effect of $\delta$ on $\offsetsize$  while fixing $\clockskew$ and $\intervalsize$.

\subsection{Analysis of Varying Message Delay ($\delta$) for the CDES}

\begin{figure}[ht]
  \centering
  \begin{subfigure}{0.4\textwidth}
    \includegraphics[width=\linewidth]{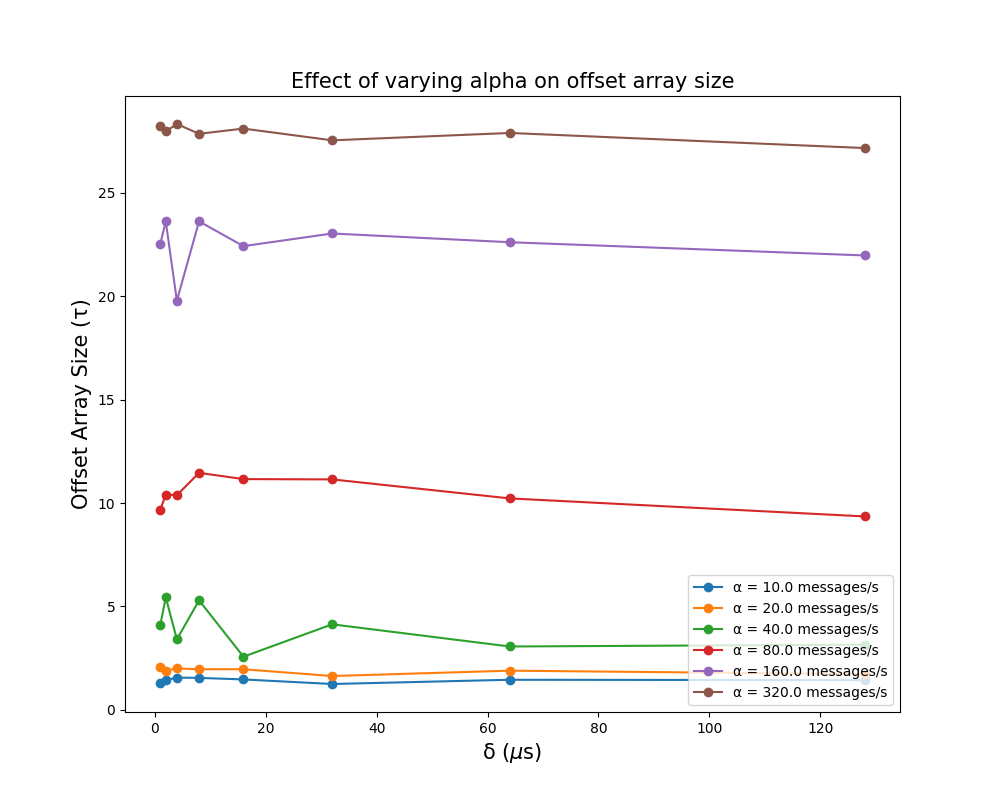}
    \caption{$n$ = 32.}
    \label{fig:effectofdelta-alpha1}
  \end{subfigure}
  \hfill
  \begin{subfigure}{0.4\textwidth}
    \includegraphics[width=\linewidth]{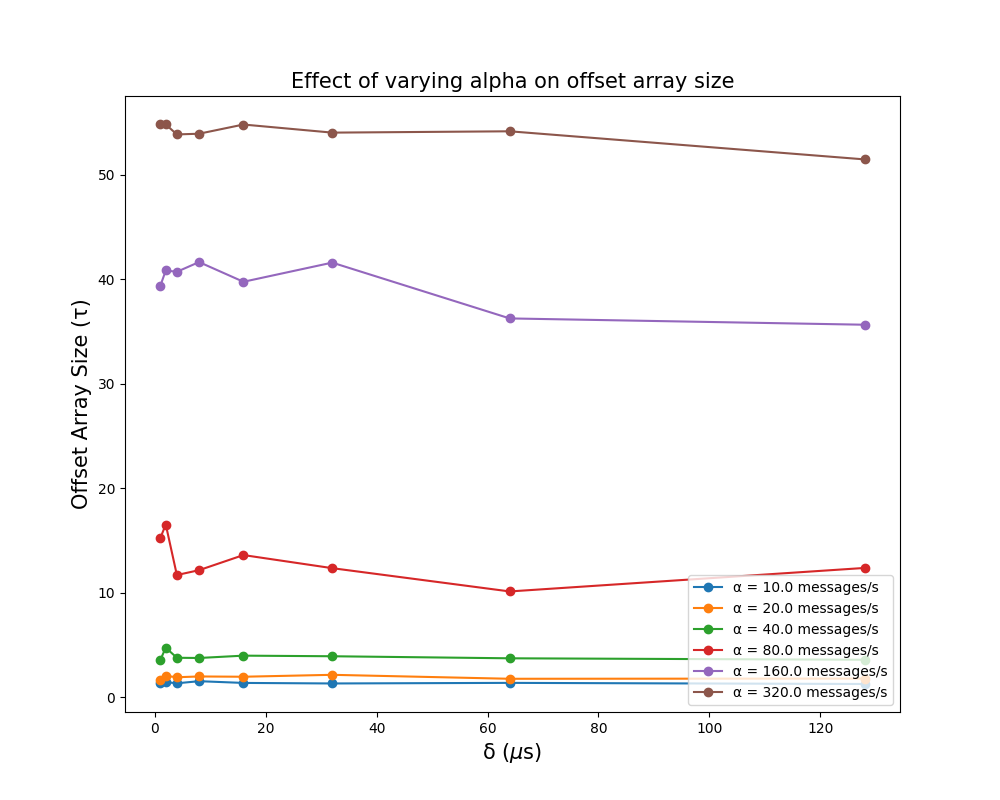}
    \caption{$n$ = 64.}
    \label{fig:effectofdelta-alpha2}
  \end{subfigure}
  \caption{Custom Simulator: $\offsetsize$ vs $\delta$ when varying $\alpha$, $\clockskew$ = 4 ms, $\intervalsize$ = 16 $\mu s$.}
  \label{fig:effectofdelta-alphaoff}
\end{figure}

Here, we observe that the value of $\delta$ has minimal effect on $\offsetsize$. Specifically, as shown in Figure \ref{fig:effectofdelta-alphaoff}, we observe that the value of $\offsetsize$ increases as the value of $\alpha$ increases. However, for a fixed value of $\alpha$, the $\offsetsize$ remains the same. When the $\clockskew$ is fixed, and the  $\intervalsize$ is fixed, the only way the $\offsetsize$ would go down is when processes would send messages that were received after the $\clockskew$ limit. Because we enforce the limit as $\delta \leq \clockskew$, this limit is never exceeded, and the $\offsetsize$ hence, remains the same. This is illustrated in Figure \ref{fig:effectofdelta-alphaoff}.

\subsection{Analysis of Varying Message Delay ($\delta$) for NS-3} 

\begin{figure}[ht]
  \centering
  \begin{subfigure}{0.4\textwidth}
    \includegraphics[width=\linewidth]{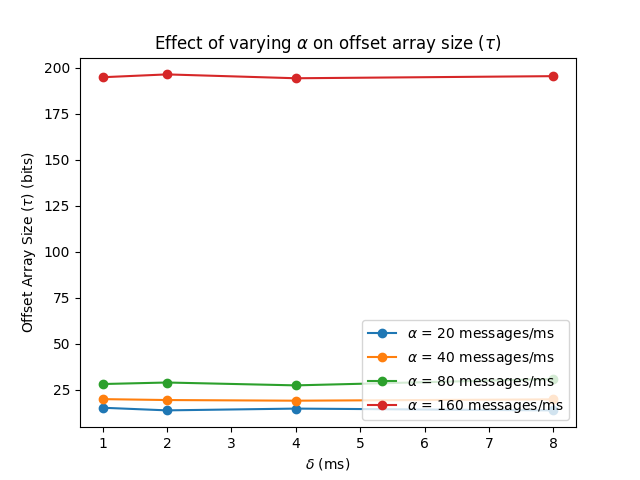}
    \caption{$n$ = 32.}
    \label{fig:ns3-effectofdelta-alpha1}
  \end{subfigure}
  \hfill
  \begin{subfigure}{0.4\textwidth}
    \includegraphics[width=\linewidth]{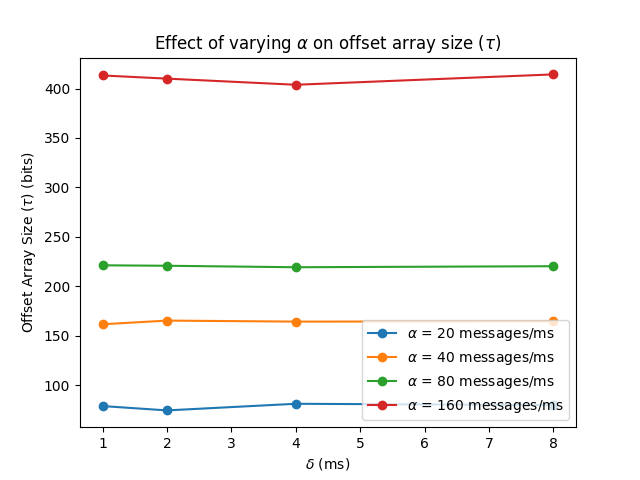}
    \caption{$n$ = 64.}
    \label{fig:ns3-effectofdelta-alpha2}
  \end{subfigure}
  \caption{NS-3 Simulator: $\offsetsize$ vs $\delta$ when varying $\alpha$, $\clockskew$ = 3 ms, $\intervalsize$ = 20 $\mu s$.}
  \label{fig:ns3-effectofdelta-alphaoff}
\end{figure}

We gather similar results in the case of NS-3 as we did in the CDES, which validates our observations. This is illustrated in Figure \ref{fig:ns3-effectofdelta-alphaoff}. We also notice that for higher values of $\alpha$, the $\offsetsize$ increases drastically. Hence, it is important to note the feasibility of the $\repcl$, mainly that it is advantageous to use it in a setting of low $\alpha$ (message rates). This is covered further in the following section.

\section{Feasibility Regions}

In this section, we review the simulations to define the notion of feasible regions. As discussed earlier, the goal of $\repcl$ is to enable the replay of a distributed computation with a small overhead. Here, we consider the case where the user identifies the expected overhead of $\repcl$ to identify scenarios under which $\repcl$ can be used to provide a \textit{perfect-replay} that meets all the requirements from Section \ref{sec:properties}. 
Since the overhead of the counters remains virtually unchanged, we only focus on the overhead of the number of offsets, i.e., the value of $\offsetsize$. 

\begin{figure}[ht]
    \centering
    \begin{subfigure}[b]{0.4\textwidth}
        \centering
        \includegraphics[width=\textwidth]{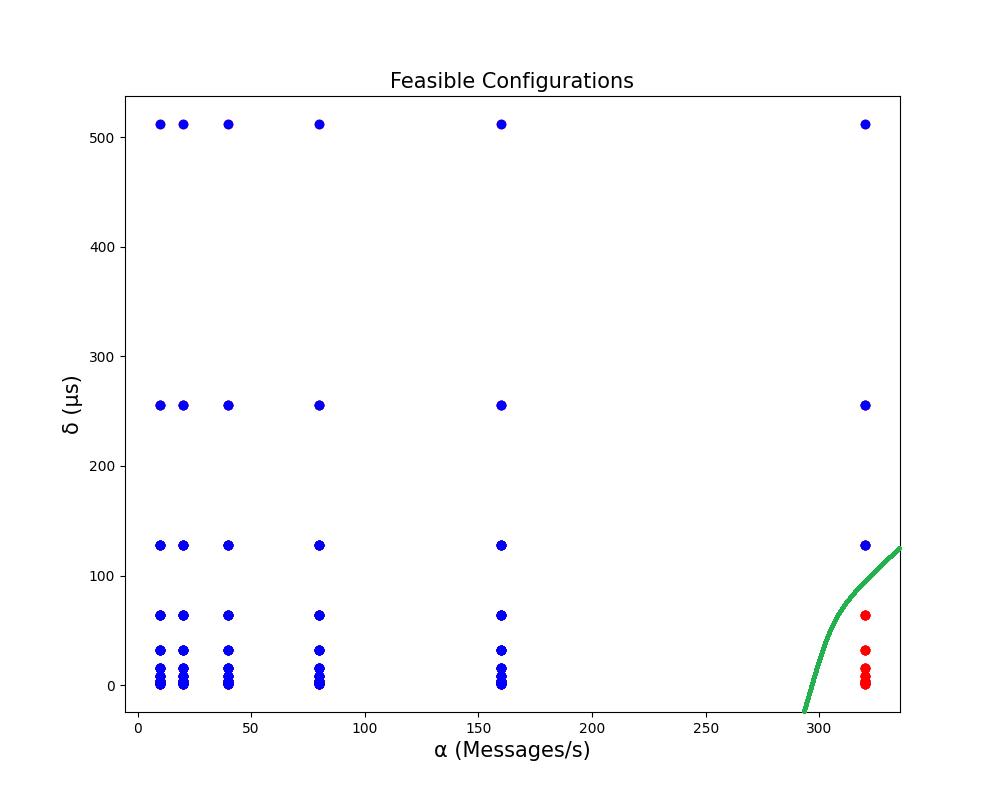}
        \caption{$\clockskew$ = 1 ms, $n$ = 32, and $\offsetsize$ = 8.}
        \label{subfig:feasible.N32.E1000}
    \end{subfigure}
    \hfill
    \begin{subfigure}[b]{0.4\textwidth}
        \centering
        \includegraphics[width=\textwidth]{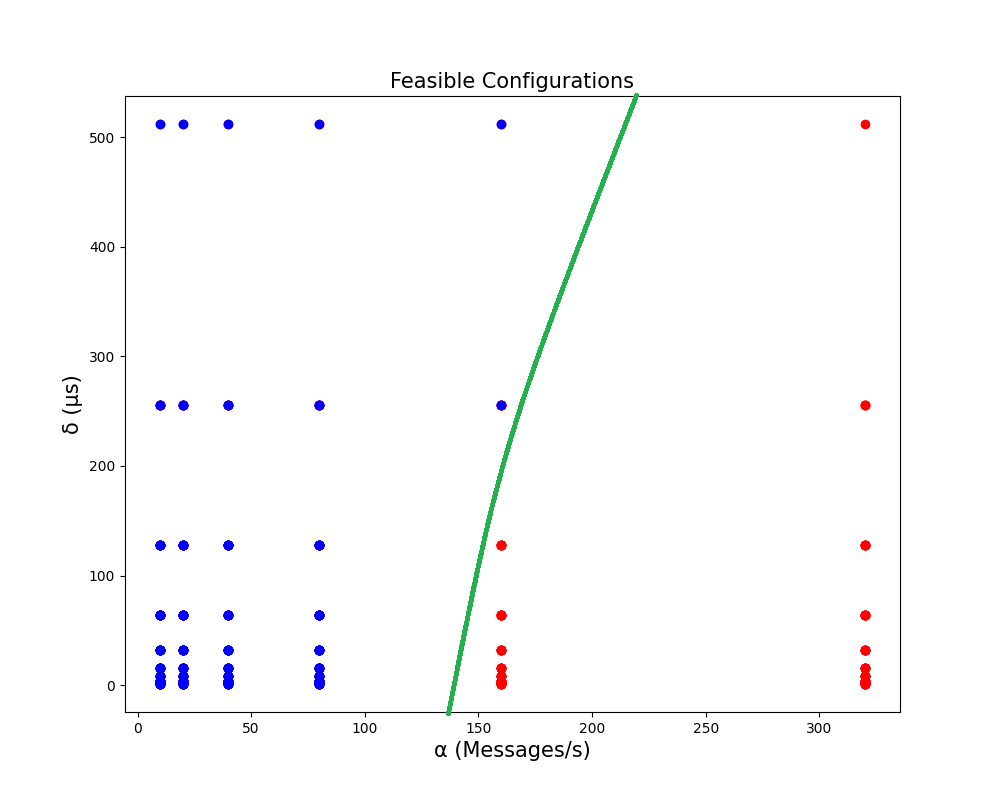}
        \caption{$\clockskew$ = 2 ms, $n$ = 32, and $\offsetsize$ = 8.}
        \label{subfig:feasible.N32.E2000}
     \end{subfigure}
    \hfill
    \begin{subfigure}[b]{0.4\textwidth}
        \centering
        \includegraphics[width=\textwidth]{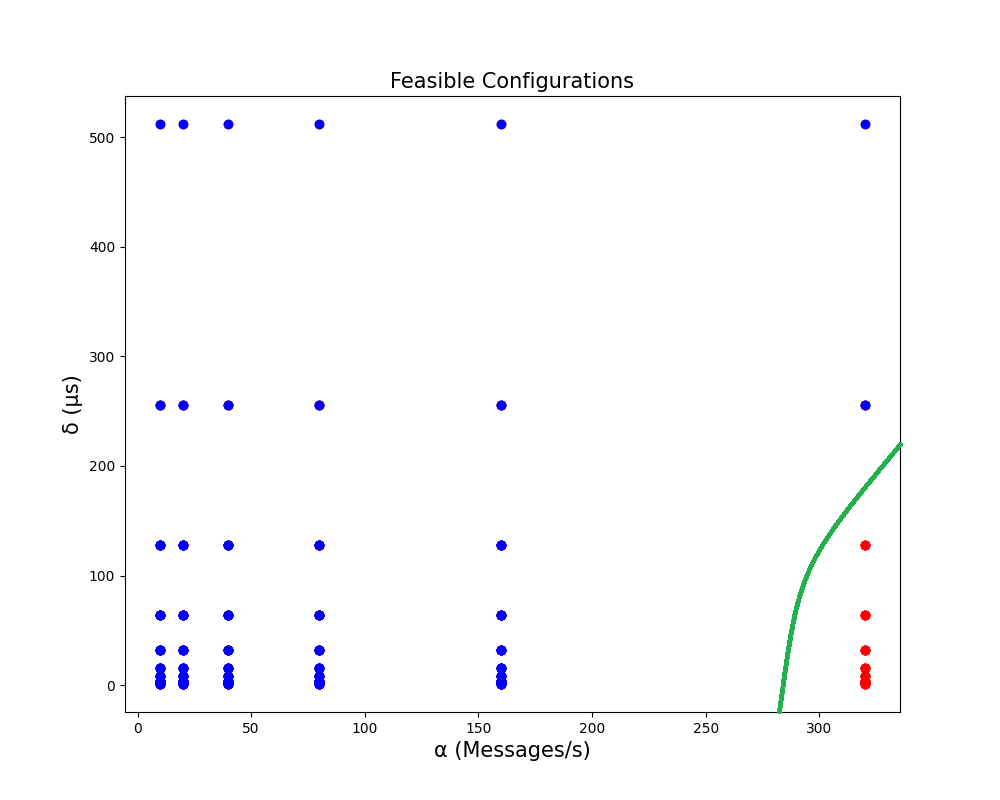}
        \caption{$\clockskew$ = 1 ms, $n$ = 64, and $\offsetsize$ = 8.}
        \label{subfig:feasible.N64.E1000}
    \end{subfigure}
    \hfill
    \begin{subfigure}[b]{0.4\textwidth}
        \centering
        \includegraphics[width=\textwidth]{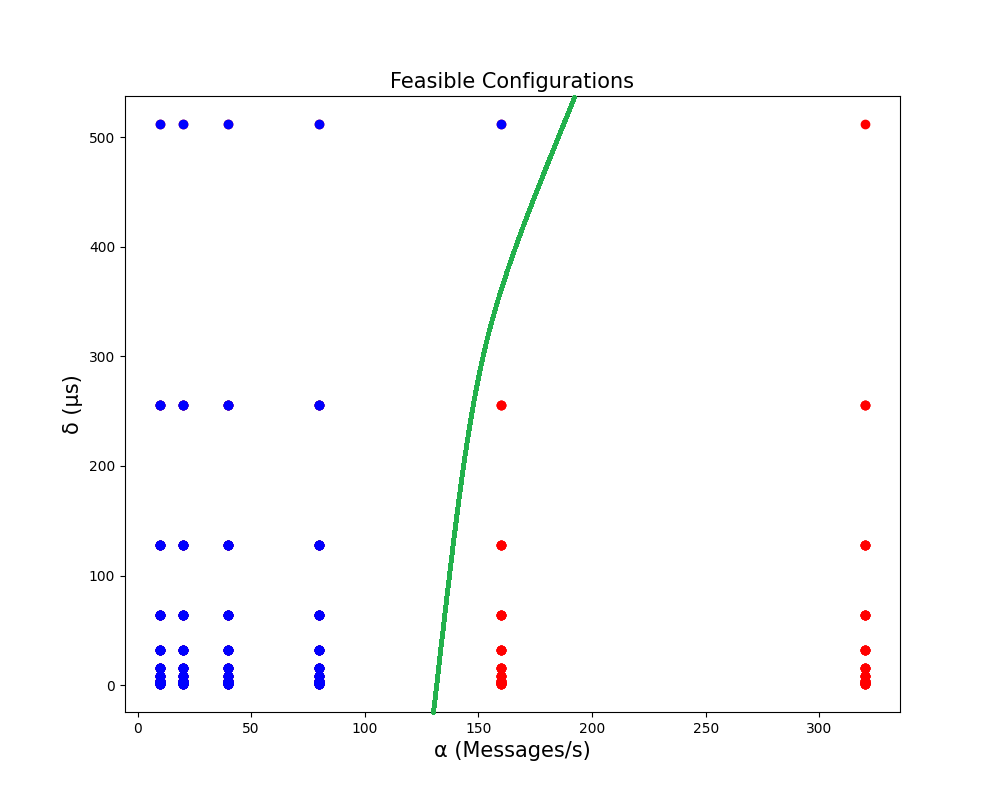}
        \caption{$\clockskew$ = 2 ms, $n$ = 64, and $\offsetsize$ = 8.}
        \label{subfig:feasible.N64.E2000}
       \end{subfigure}
    \caption{Feasibility regions for $\alpha$ and $\delta$ settings.}
    \label{fig:feasibility-regions}
\end{figure}

For $\offsetsize=8$, the feasibility regions are shown in Figure \ref{subfig:feasible.N32.E1000}. Here, the blue dots identify the data points where $\offsetsize=8$ is feasible and the red dots represent the data points where $\offsetsize=8$ is not feasible. The green line identifies the bounds where $\offsetsize=8$ is feasible. We find that the size of the feasible region remains fairly unchanged with the value of $n$. However, it shrinks when the value of $\clockskew$ is increased. This is expected based on how $\offsetsize$ changes with $\clockskew$. We note that the feasibility region only identifies the case where \textit{perfect-replay} meets all the requirements from Chapter \ref{sec:properties}. If the user needs to utilize $\repcl$ in an infeasible region, the user can obtain \textit{partial-replay}. To understand this, consider the case where the actual value of $\clockskew$ is $4ms$ but the user specifies it to be $2ms$ while constructing $\repcl$. In this case, if $e$ and $f$ are within $2ms$ then $\repcl$ will allow them to be replayed in any order. However, if $f$ occurred $3ms$ after $e$ then $e$ will always be replayed before $f$. We anticipate that even in a system where the clock skew is $4ms$, the actual clock skew at a given moment is likely to be smaller than $4ms$. This implies that the forced order between $e$ and $f$ will be quite infrequent. Hence, we anticipate that $\repcl$ will be applicable even in domains where the system parameters cause it to fall in an infeasible region, and is referred to in Section \ref{sec:discussion}.


\chapter{Visualizing Traces with \textit{RepViz}}
\label{chap:visualization}

We have now seen the various intricacies of the $\repcl$ infrastructure, and we now describe how to apply it to visualization systems. The goal of a visualization is simple, to show the user a view of the computation both from an overall system perspective and a per-process view of how the computation looked on that specific node. Various implementations of visualizers (ref. Section \ref{sec:tracer-related}) achieve this, but in all of them, the underlying timestamping algorithm enforces an order between concurrent events. We therefore, need a visualizer that allows the user to choose the order of concurrent events, and view multiple execution traces simultaneously. 

In this Chapter, we introduce \textit{RepViz}, the third infrastructure component of this work. \textit{RepViz} is a visualizer that works on top of a log generated by the $\repcl$. It takes in a $\repcl$-timestamped log, and generates a web-based visualization of the traces the algorithm generates. The user has the option to replay events that are concurrent in any order, while the other events are ordered according to the $\repcl$. Once the user has selected a replay order, a web visualization is displayed for that trace. The web visualization is under progress, but a sample representation is depicted in Figure \ref{fig:visualization}. In the following sections, we will go over the different methods that went into implementing \textit{RepViz}.

\section{Implementation}

The visualizer defines a top level component, called \textit{Tracer}. This component contains the following functions amd members:
\begin{itemize}
    \item SortEvents(): This function sorts all events according to their $\repcl$ timestamp.
    \item RunReplay(): The top level function that provides an interactive view to the user to replay events.
    \item EventList: The list of events obtained from a $\repcl$ timestamped log.
\end{itemize}

A \textit{Tracer} object contains a set of \textit{Event} objects. Each \textit{Event} object has the following properties:
\begin{itemize}
    \item EventID: The Message ID.
    \item EventType: The type of event, i.e., Send, Local or Recv.
    \item EventTime: The$\repcl$ timestamped to this event.
    \item Sender: The IP address of the sender.
    \item Receiver: The IP address of the receiver.
\end{itemize}
The Sender and Receiver fields are populated as the (Sender, Receiver) when its a Send/Local event, and as (Receiver, Sender) for a Recv event. 

Now, we will go into detail on how each of the \textit{Tracer} methods are implemented. 

\subsection{SortEvents}

The \textit{Tracer} first orders all the events according to the $\repcl$ timestamp. Events are sorted based on their $\repcl$ timestamps fed by the logs of the algorithm. The sorting algorithm sorts all timestamps by the happens-before ($\hb$) relation discussed in Chapter \ref{chap:preliminaries}. Once the ordering is set, the events are then given to the matching function. The sorting rules are described in Algorithm \ref{alg:sort_events}. The algorithm compares two $\repcl$ timestamps $t1$ and $t2$, and returns True if $t1 < t2$ and False otherwise. The other comparisons can be implicitly derived from the same algorithm.




\subsection{RunReplay}

Once the events are ordered, the \textit{Tracer} runs the replay according to the event timeline generated by the sorter. Every concurrent event pool is given to the user as a choice to replay any one of the outstanding events waiting to be replayed. Once an event has been replayed, that event is removed from the replay pool. This follows Algorithm \ref{alg:replay-events}.

\section{User View}

In the current iteration of \textit{RepViz}, the prototype is run on an ASCII terminal. Here is a brief run output of a sample trace snippet generated through an NS-3 simulation:

\begin{lstlisting}
[(EventID=1, EventType=SEND, EventTime=[(NodeId=10.1.1.3, HLC=21, Offsets=[-15, -15, 0, -15, -15], Counters=0)], Sender=10.1.1.3, Receiver=10.1.1.4)]
[(EventID=2, EventType=SEND, EventTime=[(NodeId=10.1.1.1, HLC=42, Offsets=[0, -15, -15, -15, -15], Counters=0)], Sender=10.1.1.1, Receiver=10.1.1.2)]
[(EventID=3, EventType=SEND, EventTime=[(NodeId=10.1.1.4, HLC=44, Offsets=[-15, -15, -15, 0, -15], Counters=0)], Sender=10.1.1.4, Receiver=10.1.1.5)]
[(EventID=4, EventType=SEND, EventTime=[(NodeId=10.1.1.2, HLC=55, Offsets=[-15, 0, -15, -15, -15], Counters=0)], Sender=10.1.1.2, Receiver=10.1.1.3)]
[(EventID=4, EventType=RECV, EventTime=[(NodeId=10.1.1.3, HLC=55, Offsets=[-15, 0, 0, -15, -15], Counters=1)], Sender=10.1.1.3, Receiver=10.1.1.2)]
[(EventID=5, EventType=SEND, EventTime=[(NodeId=10.1.1.1, HLC=57, Offsets=[0, -15, -15, -15, -15], Counters=0)], Sender=10.1.1.1, Receiver=10.1.1.2)]
[(EventID=6, EventType=SEND, EventTime=[(NodeId=10.1.1.3, HLC=59, Offsets=[-15, 4, 0, -15, -15], Counters=0)], Sender=10.1.1.3, Receiver=10.1.1.4)]
[(EventID=7, EventType=SEND, EventTime=[(NodeId=10.1.1.5, HLC=61, Offsets=[-15, -15, -15, -15, 0], Counters=0)], Sender=10.1.1.5, Receiver=10.1.1.1)]
[(EventID=7, EventType=RECV, EventTime=[(NodeId=10.1.1.1, HLC=61, Offsets=[0, -15, -15, -15, 0], Counters=0)], Sender=10.1.1.1, Receiver=10.1.1.5)]
[(EventID=8, EventType=SEND, EventTime=[(NodeId=10.1.1.1, HLC=61, Offsets=[0, -15, -15, -15, 0], Counters=1)], Sender=10.1.1.1, Receiver=10.1.1.2)]
Concurrent events detected!
0. [(EventID=9, EventType=SEND, EventTime=[(NodeId=10.1.1.2, HLC=62, Offsets=[-15, 0, -15, -15, -15], Counters=0)], Sender=10.1.1.2, Receiver=10.1.1.3)]
1. [(EventID=10, EventType=SEND, EventTime=[(NodeId=10.1.1.5, HLC=62, Offsets=[-15, -15, -15, -15, 0], Counters=0)], Sender=10.1.1.5, Receiver=10.1.1.1)]
2. [(EventID=2, EventType=RECV, EventTime=[(NodeId=10.1.1.2, HLC=62, Offsets=[0, 0, -15, -15, -15], Counters=0)], Sender=10.1.1.2, Receiver=10.1.1.1)]
Please choose the event to replay: 0
[(EventID=9, EventType=SEND, EventTime=[(NodeId=10.1.1.2, HLC=62, Offsets=[-15, 0, -15, -15, -15], Counters=0)], Sender=10.1.1.2, Receiver=10.1.1.3)]
Please choose the event to replay: 2
[(EventID=2, EventType=RECV, EventTime=[(NodeId=10.1.1.2, HLC=62, Offsets=[-15, 0, -15, -15, -15], Counters=1)], Sender=10.1.1.2, Receiver=10.1.1.1)]
Please choose the event to replay: 1
[(EventID=10, EventType=SEND, EventTime=[(NodeId=10.1.1.5, HLC=62, Offsets=[-15, -15, -15, -15, 0], Counters=0)], Sender=10.1.1.5, Receiver=10.1.1.1)]
\end{lstlisting}

\begin{figure}[ht]
  \centering
  \begin{subfigure}{0.5\textwidth}
    \includegraphics[width=\linewidth]{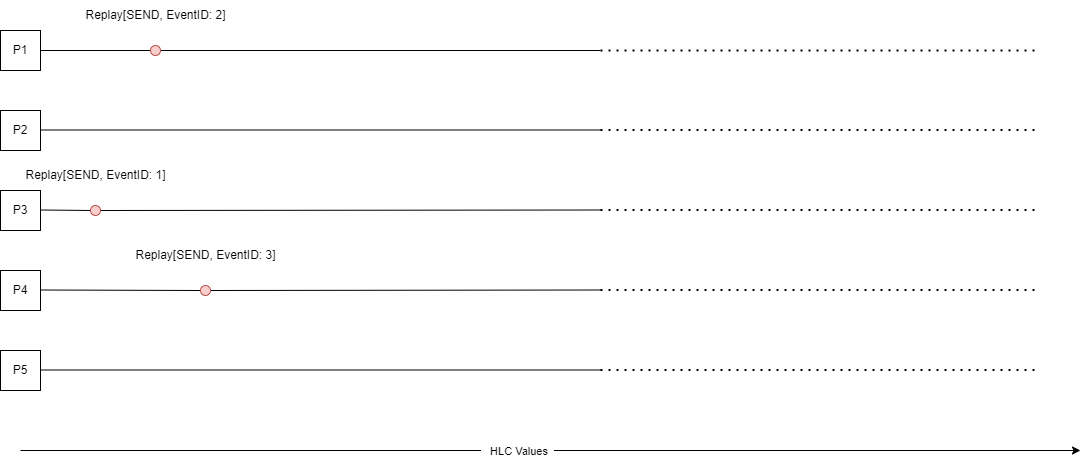}
    \caption{No concurrency conflicts, push the right arrow key.}
    \label{fig:tracer1}
  \end{subfigure}
  \hfill
  \begin{subfigure}{0.5\textwidth}
    \includegraphics[width=\linewidth]{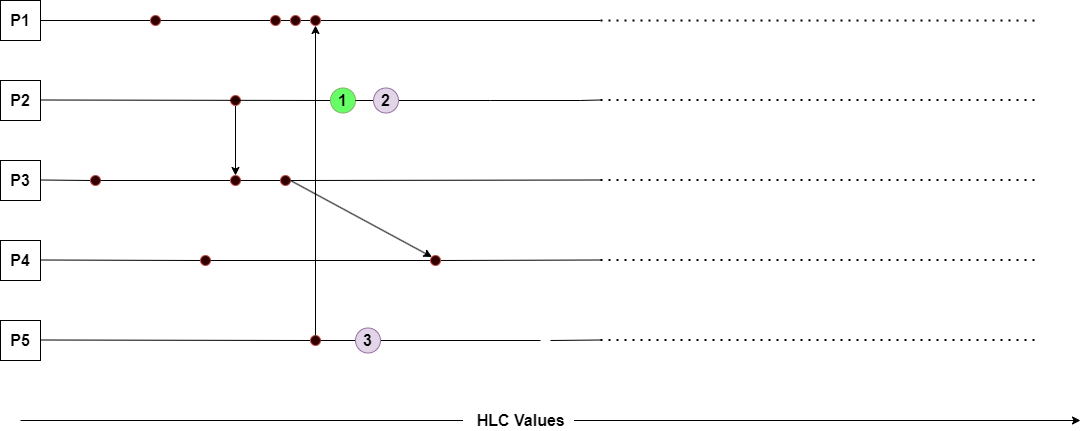}
    \caption{Concurrency conflicts detected, replay event 1 first.}
    \label{fig:tracer2}
  \end{subfigure}
  \hfill
  \begin{subfigure}{0.5\textwidth}
    \includegraphics[width=\linewidth]{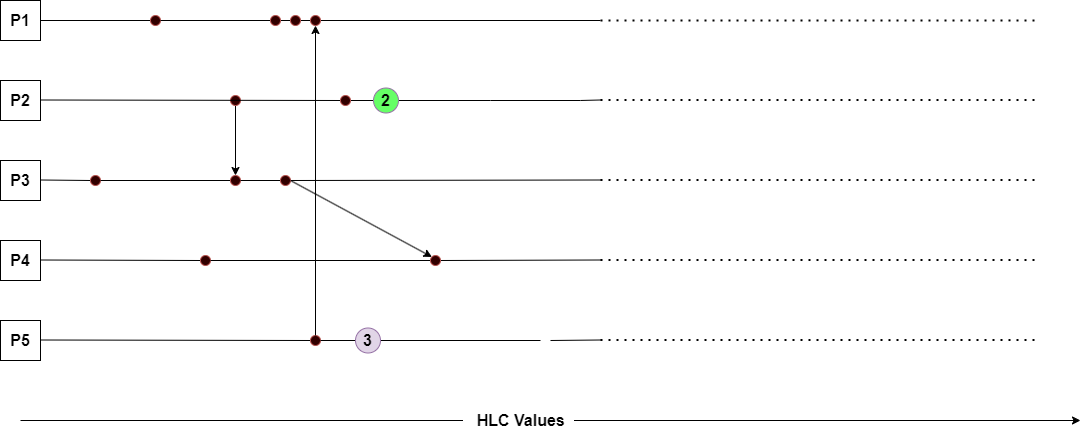}
    \caption{Event 1 replayed, replay event 2 next.}
    \label{fig:tracer3}
  \end{subfigure}
  \hfill
  \begin{subfigure}{0.5\textwidth}
    \includegraphics[width=\linewidth]{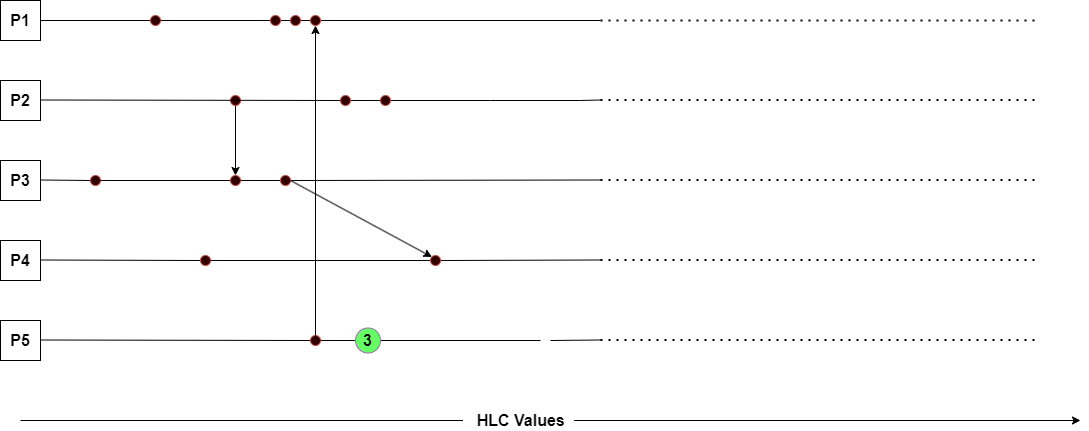}
    \caption{Event 2 replayed, replay event 3 next.}
    \label{fig:tracer4}
  \end{subfigure}
  \caption{Sample Visualization: Here the user uses arrow keys to replay events that do not have concurrency conflicts, and then uses number keys to input which events to replay in which order.}
  \label{fig:visualization}
\end{figure}

The above log would transform to the visualization on a WebUI as shown in Figure \ref{fig:visualization}. In Figure \ref{fig:tracer1}, the user does not have a choice in the replay, as all events are ordered with the $\hb$ relation detailed in Section \ref{sec:compare} in the absence of concurrency. The user simply taps the right arrow key to move forward in the replay. At some point in the replay, the user encounters concurrent events, depicted in Figure \ref{fig:tracer2}. Here, the user elects to replay event 1 by providing an input of 1 through the keyboard. The event marked 1 is added to the replay. Next the user has to choose between events 2 and 3, depicted in Figure \ref{fig:tracer3}. The user chooses event 2, and the last event left to replay is event 3, which is replayed in Figure \ref{fig:tracer4}. With this visualization, it is easy for a user to try different combinations of replay and analyse how parameters change with the event order chosen. The visualization can also generate exhaustive logs of all possible replays should the user require it.  

\chapter{Related Work and Discussion}
\label{chap:related}

\section{Clocks in Distributed Systems}

Logical clocks were proposed in 1978 by Leslie Lamport \cite{Lamport78CACM} to trace the ordering of events in a distributed system. Vector time was designed independently by multiple researchers \cite{nagel1996vampir}\cite{mattern1988virtual}\cite{schmuck1988use}, and they proposed the idea of representing time in a distributed system as a set of n-dimensional non-negative integer vectors. According to \cite{kshemkalyani2011distributed}, Vector clocks are defined by three properties: Isomorphism, Strong consistency, and Event Counting. Isomorphism suggests that if two events $x$ and $y$ have timestamps $vh$ and $vk$, respectively, then $x \longrightarrow y \Longleftrightarrow vh < vk$. Here, $\longrightarrow$ implies a partial ordering between a set of events. Strong consistency implies that by examining the vector timestamp of two events, we can determine the causal relationship between the two events. Event counting suggests that if $d$ is always 1 in the rule $R1$, then the $i^{th}$ component of the vector clock at process $p_i, vt_i[i]$, denotes the number of events that have occurred at $p_i$ until that instant. 

There have been several prior implementations of vector clocks including Singhal-Kshemkalyani’s differential technique \cite{singhal1992efficient} and Fowler-Zwaenepoel direct dependency technique \cite{fowler1990causal}. While vector clocks are upper bounded by $O(n)$ complexity in terms of both time and memory complexity, the different implementations of the past have tried to reduce this complexity and generate more efficient representations, with some success. Singham-Kshemkalyani's differential technique relied on piggybacking using the last sent and last update, without updating every vector clock. This method relies on the assumption that even though the number of processes is large, only a few key processes in a system would interact frequently by passing messages. A benefit of this method is that it cuts down storage overhead at each process to $O(n)$. However, this method doesn't make a substantial contribution to reducing the time complexity incurred when updating the vector clock, as it relies on piggybacking to work. 

Fowler-Zwaenepoel's direct-dependency technique cuts down storage complexity again by reducing the message size during transmission by transmitting only a scalar value in the messages. Here, a process only maintains information regarding direct dependencies on other processes. The downside of this method is that it has a high computational overhead as it has to trace dependencies and update the vector clock, especially in systems where a few key processes may have a large number of events. 

Clock synchronization using Network Time Protocol (NTP) uses the Offset Delay Estimation method to ensure physical clocks are synchronized across the internet. Clock offsets and delays are calculated, and timestamps are issued between different machines within a system accordingly. The system then attempts to establish a causal relationship by using the corrected timestamps. This, however, can be computationally expensive and is open to error, as the delay estimation may not always be accurate and result in a violation of the causal relationship between processes issued by different machines. 

One existing limitation between vector clocks representing logical time and physical clock synchronization is the difficulty in reconciling one with the other. To overcome this challenge, Hybrid Logical clocks were introduced by Kulkarni et al. \cite{kulkarni2014logical} to capture the causality relationship of a logical clock with the characteristics of a physical clock embedded into it. Another variant of the hybrid clock is the Hybrid Vector Clock \cite{demirbas2013beyond}, \cite{yingchareonthawornchai2018analysis}, which, unlike the Hybrid Logical Clock, can provide all possible/potential consistent snapshots for a given time. For this experiment, we use the Hybrid Vector Clock design presented in Yingchareonthawornchai et al. as it provides desirable characteristics to build our visualization framework.

\section{Visualizing Traces}
\label{sec:tracer-related}

Mattern \cite{mattern1988virtual} talks about how distributed systems use the concept of global state to communicate information, and the need to characterize this global state. They talk about how a process can only approximate the global view of the system, and no process can have, at any given instant, a consistent view of the global state. To verify a distributed system, the author provides a comparison between three key approaches: simulating a synchronous distributed system given an asynchronous system, simulating a common clock or simulating a global state. They highlight the need of a vector clock system to provide a consistent snapshot of the global state, as each process having a clock that stores only its own state is not enough to describe the global state of the computation. 

PARAVER \cite{pillet1995paraver} uses the PVM message passing library to analyze traces generated from a computation. PVM primarily uses parallel message passing, and PARAVER analyses these parallel traces using data analytics and provides a graphical description of the analysis. This was one of the earliest visualization works on distributed systems simulated only for parallel traces. It used a parser to parse through the logs of the PVM-generated traces and analyze CPU activity, communications, and user events. This however required the addition of functionality to the PVM itself and was not generalized to any distributed system interface. It also did not provide generic support and incurred a larger overhead to profile system resources while computation went on.

VAMPIR \cite{nagel1996vampir} provided analysis of MPI programs by generating timeline traces by profiling MPI applications. It used different visualization metrics to show whether processes were still active or not. It also provided views of system activities and aggregated statistics about the system itself. However, it was made specifically only for MPI applications and added to the profiling interfaces of MPI.

D3S \cite{liu2008d3s} allowed developers to specify predicates on distributed properties of the system. These predicates can vary depending on what consistency checks one would require on the distributed system. They modeled the tracing as a consistency checker and generated traces of predicate evaluation. The predicates are injected dynamically at compile time into the system and are evaluated based on the customization provided by the user. However, we believe this approach would add overhead to running the distributed computation due to complex predicate checking.

Zinsight \cite{de2010zinsight} provides hierarchies of tasks and provides aggregated metrics to show timeline visualizations of events. It also provides users with changing the granularity of the metric they want to see with sequences of computations per process. 

Trumper et al. \cite{trumper2010understanding} present a dynamic analysis tool that uses boundary tracing and post-processing to analyze system behavior through a distributed computation. These are task-based visualizations, where tasks are mapped to memory resources. However, this may not always be the case where processes share the same memory, as in the case of OpenMP-based infrastructures.

Dapper \cite{sigelman2010dapper} is Google's tracing software for distributed systems where they provided low overhead, application-level transparency, and scalability. Dapper uses annotations and \textit{spans} to generate traces through RPCs. However, the authors mention that Dapper cannot correctly point to causal history, as it uses annotations in non-standard control primitives, that may mislead the causality calculations. Our approach would overcome this, as causality is enforced through a lattice of clocks, rather than the events itself.

Isaacs et al. \cite{isaacs2014state} provides a comprehensive survey of different distributed monitoring and tracing tools in the past decade, providing detailed descriptions and categorizations based on task parallelism, causality information, and so on.

Isaacs, Bremer et al. \cite{isaacs2014combing} design a trace visualization system purely relying on logical clocks and then transposing those clocks back to real-time clocks in the visualization. Processes are also clustered based on logical behavior. However, this would incur more overhead than our solution and may cause conflicts in enforcing causality, due to the usage of a standard logical clock.

Verdi \cite{wilcox2015verdi} provides developers with choosing the fault system to diagnose, and verify the implementation of the system. This is a formal verification system where it provides the developer with an idealized fault model, and once this is verified, it applies the correctness to a more realistic fault model.

ShiViz \cite{beschastnikh2016debugging} uses vector clocks in generating distributed system traces using happens-before relationships. By using vector clocks, it provides a verifiable and accurate notion of causality. However, since it uses traditional vector clocks, it uses a higher complexity than our proposed model.

\section{Discussion}
\label{sec:discussion}

In Section \ref{chap:simulation}, we identified feasible regions for the given permissible overhead for $\repcl$. Thus, the natural question is: \textit{what can a user do if the given system parameters fall into the infeasible region?} Here, observe that $\clockskew$ provides one way to reduce the overhead if we accept some imperfect replay. To explain this, consider the case where we are using a system with clock skew to be $\clockskew_a$. If the user implements $\repcl$ with  $\clockskew < \clockskew_a$ then the resulting replay will still satisfy requirements 1 and 2 (cf. Section \ref{sec:properties}). Requirement 3 will be satisfied with $\epsilontwo=\clockskew-\intervalsize$.

\begin{figure}
    \centering
    \includegraphics[width=0.5\linewidth]{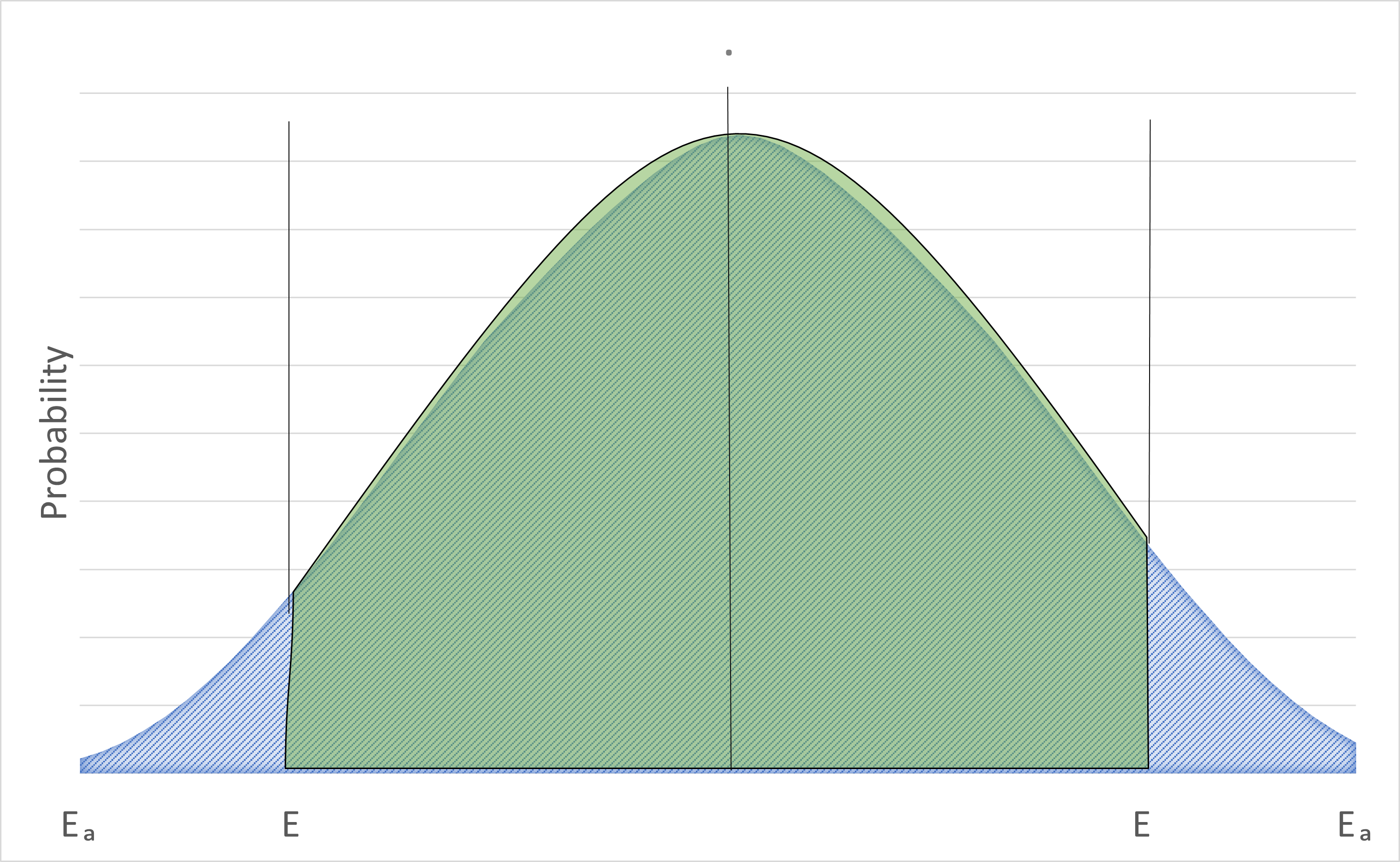}
    \caption{Effect of using $\clockskew$ instead of $\clockskew_a$ in $\repcl$.}
    \label{fig:epsvsprob}
\end{figure}

Looking at this situation closely, we observe that the clock skew between two processes follows a structure shown in Figure \ref{fig:epsvsprob}. Specifically, at a given instance, the clocks of two processes $j$ and $k$ differ by some amount that is less than $\clockskew_a$. However, the actual clock difference at a fixed point in time (that is not visible to either $j$ and $k$) is often less than $\clockskew_a$. Hence, if $e$ and $f$ occurred at the same global time, the probability that the respective clocks differed by $\clockskew$ depends upon the area of the shaded part (cf. Figure \ref{fig:epsvsprob}). In this case, $e$ and $f$ would still be replayed in arbitrary order. Only if the clocks fall in the non-shaded area then the replay will force an order between $e$ and $f$. In other words, even if the system parameters fall in the infeasible region, it would be possible to use $\repcl$ that provides a valid replay. It is just that it will not be able to reproduce all possible replays. 

\chapter{Conclusion and Future Work}
\label{chap:conclusion}

In this paper, we focused on the problem of replay clocks in systems where clocks are synchronized to be within $\clockskew$. The purpose of these clocks is to reproduce a distributed computation with all its certainties and uncertainties. By certainty, we mean that if event $e$ must have happened before $f$ then the replay must ensure that $e$ is replayed before $f$. Specifically, this required that if $e$ happened before $f$ (as defined in \cite{Lamport78CACM}) or $f$ occurred $\epsilonone\approx\clockskew$ time after $e$ then $e$ must occur before $f$.
And, by uncertainty, we mean that if $e$ and $f$ could occur in any order then the replay should not force an order between them. Specifically, if $e||f$ (as defined in \cite{Lamport78CACM}) and $e$ and $f$ occurred within time $\epsilontwo\approx\clockskew$ then the replay permits them to be replayed in any order. We presented $\repcl$ to solve the replay problem with $\epsilonone=\clockskew+\intervalsize$ and $\epsilontwo=\clockskew-\intervalsize$, where $\intervalsize$ is a parameter to $\repcl$.  
We analyzed $\repcl$ for various system parameters (clock skew ($\clockskew$), message rate ($\alpha$), message delay ($\delta$)). We find that for various system parameters, the size of $\repcl$ and the overhead to create timestamps and/or compare them is small. For the purpose of replay, $\repcl$ provides several advantages over existing approaches. For example, unlike logical clocks, they do not force certain unneeded event ordering. They have a significantly lower overhead compared to vector clocks. Also, they do not generate illegitimate replays that can occur with the user of vector clocks. The overhead of $\repcl.j$ depends upon the number of processes that communicate with $j$ (directly or indirectly) in $\clockskew$ window. This is different from the case in vector clocks where the overhead is always proportional to the number of processes in the system. 

With the design of the $\repcl$, we ensured that the clock size is not a leading factor in slowing down computation. The $\repcl$ is a non-invasive method to ensure that causality is maintained in the presence of skewing clocks, and this is particularly useful in various applications. To facilitate the ease of development with this clock, we provide an API and a sample implementation of the API in NS-3, a widely used distributed network simulator. We illustrate with the help of NS-3, the various invariants our clock provides, such as the size scaling while varying various parameters. We also identify feasibility regions that would provide perfect replay through the clock. For systems out of this feasible region, we additionally provide techniques to approximate replay that is acceptable. 

We have utilized $\repcl$ to enable users to visualize a distributed computation using our tool, \textit{RepViz}. The goal of the visualization is to allow the user to identify an event $f$ where a failure occurred. Then, they can use a replay of events just preceding $f$ to determine whether the error would go away. If it does, it would imply that it is a synchronization error. Likewise, a user can replay some portion of the computation. Since the replay of events may occur in a different order, it will help identify potential synchronization errors. 

$\repcl$ is designed mainly for offline analysis, where the event data is stored during execution and analyzed at a later time. However, $\repcl$ can be used for run-time monitoring/analysis as well if the data related to the timestamps is sent to a monitor, that monitor could analyze it for potential properties of interest if the analysis can be done \textit{quickly}. However, a key challenge in this context will be whether the run-time monitors can keep up with the execution of the system. There are several future directions for $\repcl$. If the size of $\repcl$ needed for perfect replay is too large, the user can reduce the size of $\repcl$ by choosing a lower value of $\clockskew$. In this case, the resulting replay will force some ordering between concurrent events. One of the future works is to identify the effect of reducing $\clockskew$ in this manner. 

Another potential future extension would be the ability to evaluate different veins of execution for different properties. Currently, it is difficult to compare and contrast different traces of execution for a specific invariant. With the help of \textit{RepViz}, users can identify key characteristics on how data is moved between processes, and if there is an efficient way to coordinate movement. Another possible characteristic that could be measured is resource utilization and how it is affected by different veins of execution. Specifically, we aim to answer the question: Are there stark differences when an event is replayed first before another based on the amount of work needed to perform that event? This would give better methodologies to evaluate data movement operations.

Furthermore, we intend to expand this clock structure to beyond 64 processes. A potential avenue to do this is to implement it on a hierarchical structure. If a network is structured as a network of switches, with each switch connected to a cluster of nodes, we can implement a replay clock for each level independently. All we would need is a mechanism to merge the clocks coming in from the cluster to the switch and have the switch relay clock information from its cluster to the other clustered nodes on the network. 

%
%
%
%
%
\makebibliographypage
\bibliography{refs}
\bibliographystyle{ieeetr}
%
%
%
%
%
%
%
%
%
\end{document}